\documentclass[epj,final,amsmath,amssymb]{svjour}

\def\<{{<}}
\def\>{{>}}

\usepackage{graphicx}

\input{epsf}

\begin{document}

\title{Quantum chaos and random matrix theory
for fidelity decay \\ in quantum computations
with static imperfections}

\author{Klaus M. Frahm,  Robert Fleckinger and Dima L. Shepelyansky}

\institute{Laboratoire de Physique Th\'eorique, UMR 5152 du CNRS, 
Universit\'e Paul Sabatier, 31062 Toulouse Cedex 4, France}

\titlerunning{Quantum chaos and random matrix theory for fidelity decay}
\authorrunning{K.M.Frahm,  R.Fleckinger and D.L.Shepelyansky}

\date{Received: December 13, 2003}

\abstract{
We determine the universal law for fidelity decay
in quantum computations of complex dynamics in presence
of internal static imperfections in a quantum computer. 
Our approach is based on random matrix theory applied to
quantum computations in presence of imperfections.
The theoretical predictions are tested and confirmed 
in extensive numerical simulations of a quantum algorithm for
quantum chaos in the dynamical tent map with up to 18 qubits.
The theory developed determines the time scales for
reliable quantum computations in absence of 
the quantum error correction codes. These time scales are related 
to the Heisenberg time, the Thouless time, and the decay time given by 
Fermi's golden rule which are well known in the context of mesoscopic 
systems. 
The comparison is presented for static imperfection effects and random 
errors in quantum gates. A new convenient
method for the quantum computation
of the coarse-grained Wigner function is also proposed.
}

\maketitle
\section{Introduction}
\label{sec1}

Recently a great deal of attention has been attracted to
the problem of quantum computation (see {\it e.g.} \cite{josza,steane,chuang}).
A quantum computer is viewed as a system of qubits.
Each qubit can be considered as a two-level quantum system,
{\it e.g.} one-half spin in a magnetic field. 
For $n_q$ qubits the whole system is characterized by a finite - dimensional
Hilbert space with $N\;=\;2^{n_q}$ quantum states. It has been shown
that all unitary operations in this space can be realized
with elementary quantum gates which include one-qubit rotations $B^{(1)}$
and two-qubit controlled operations, {\it $\;$ e.g. $\;$} controlled-NOT
gate $C^{(N)}$ or controlled phase-shift gates $B^{(2)}(\phi)$
(see {\it e.g.} \cite{chuang,divi}). The gates $C^{(N)}$ and $B^{(2)}(\phi)$
assume that the interaction between qubits can be switched on and off in a 
controllable way with sufficiently high accuracy. 
Various computational algorithms in the space $N$
can be represented as a sequence of elementary gates.
A general unitary operation (unitary matrix) in this space requires
an exponential  ( in $n_q$) number of elementary gates. However, there are
important examples of algorithms for which the quantum computation
can be performed with  a number of operations (gates)
much smaller than with the classical algorithms.
The most famous is the Shor algorithm for factorization of 
integers with $n_q$ digits
which on a quantum computer can be performed with
$O(n_q^3)$ gates contrary to an exponential
number of operations   required for any known classical algorithm
\cite{shor}. Another example is the Grover algorithm 
for a search of unstructured database
which has a quadratic speedup comparing to any classical algorithm
\cite{grover}. 

A quantum computation can be much faster
than a classical one due the massive parallelism of many-body
quantum mechanics 
since any step of a quantum evolution is a 
multiplication of 
a vector by a unitary matrix. A very important example
is the quantum Fourier transform (QFT) which can be performed for a vector
of size $N=2^{n_q}$ with $O({n_q}^2)$ gates instead
of $O(n_q 2^{n_q})$ classical operations required for
the fast Fourier transform (FFT) (see {\it e.g.} \cite{josza,chuang}). 
With the help of QFT the quantum evolution of certain 
many-body quantum systems
can be performed in a polynomial number of gates \cite{lloyd,ortiz}.
Another example can be found in the evolution of quantum dynamical systems
which are chaotic in the classical limit
(see {\it e.g.} \cite{chirikov,izrailev}). Such systems 
are described by chaotic quantum  maps and include
the quantum baker map \cite{schack}, the 
quantum kicked rotator \cite{georgeotkr}, the quantum saw-tooth map 
\cite{benentist} and the quantum double-well map \cite{well}.
For them a map iteration can be performed for $N$-size vector
in $O({n_q}^2)$ or $O({n_q}^3)$ gates while a classical 
algorithm would need $O(n_q 2^{n_q})$ operations. 
This however does not necessary lead to an exponential gain
since the final step with extraction of information by measurements
also should be taken into account. Thus, for example,
the quantum simulation of the Anderson metal-insulator
transition gives only a quadratic speedup
even if each step of quantum evolution is performed in
a polynomial number of gates \cite{pomeransky}.
Among other algorithms, let us refer to the quantum
computation of  classical chaotic dynamics
where some new information can be obtained efficiently
\cite{cat,stratt}.

The main obstacle to experimental implementation of a quantum computer
is believed to be decoherence induced by unavoidable couplings
to external world (see {\it e.g.} \cite{zurek}). 
However, even if we imagine that there are no external couplings
 there still remains internal static imperfections
inside a quantum computer. These static imperfections generate
residual couplings between qubits and variation of energy level-spacing 
from one qubit to another. As it was shown in \cite{georgeot}
such imperfections lead to emergence of many-body quantum chaos
in a quantum computer hardware if a coupling strength exceeds 
a quantum chaos threshold. 
In a realistic quantum computer this threshold drops only 
inversely proportionally to the number of qubits $n_q$ 
while the  energy spacing between nearby levels drops
exponentially with $n_q$.   The dependence of this threshold 
on quantum computer parameters was studied analytically and numerically
by different groups \cite{georgeot,flambaum,berman,benentieu,braun}
The time scales for onset
of quantum chaos were also determined.

It is of primary importance to understand how effects of
external decoherence and internal static imperfections
affect the accuracy of quantum computations. A very convenient
characteristic which allows to analyze these effects is the
fidelity $f$ of quantum computation. It is defined as
$f(t)=|<\psi_{\varepsilon}(t)|\psi (t)>|^2$
where $|\psi (t)>$ is the quantum state at time $t$ computed with
perfect (or ideal) gates, while
$|\psi_{\varepsilon}(t)>$ is the quantum state at time $t$ 
computed with imperfect gates characterized by an imperfection strength
$\varepsilon$. If the fidelity is close to unity then
a quantum computation with imperfections is close to the ideal one
while if $f$ is significantly smaller than  1 then
the computation gives, generally, wrong results.

At first the fidelity was used to characterize the effects of
perturbation on quantum evolution in the regime of quantum chaos
\cite{peres}. Indeed, for the classical chaotic dynamics
the small errors grow exponentially with time while 
for the quantum evolution 
in the regime of quantum chaos small quantum errors 
only weakly affect the dynamics. For example, the time reversibility
is broken by small errors for classical chaotic dynamics 
while it is preserved for the corresponding quantum dynamics
in presence of small quantum errors \cite{dls1983,casati1986}.
In the context of quantum computation the qualitative difference between
classical and quantum errors is analyzed in \cite{cat}.
Recently, the interest to the fidelity decay induced by 
perturbations of dynamics in the regime of quantum chaos
has been renewed 
\cite{pastawski,beenakker,veble,como,prosen,cohen,cerruti,adamov}.
It has been shown that the rate $\Gamma$ of exponential decrease of $f$
is given by the Fermi golden rule for small perturbations
while for sufficiently strong perturbations
the decay rate is determined by the Kolmogorov-Sinai entropy
related to the Lyapunov exponent of classical chaotic dynamics
\cite{beenakker,veble}. For small perturbations the fidelity
decay can be expressed with the help of  correlation function
of quantum dynamics that allows to understand
various peculiarities of the decay \cite{prosen}. 

Until recently the fidelity decay and accuracy of quantum computations
have been mainly analyzed for the case of random noise errors in the quantum 
gates \cite{paz,cat,benentist,well,wavelet,bettelli}. 
Quite naturally in this case
the rate of fidelity decay is proportional
to the square of error amplitude $\varepsilon$ 
($\Gamma \propto \varepsilon^2$). Indeed, a random error of amplitude
$\varepsilon$ transfers a probability of order  $\varepsilon^2$
from the ideal state to all other states and as a result
the fidelity remains close to unity (within {\it e.g.}
10\% accuracy) during a time scale
$t_f \sim 1/(\varepsilon^2 n_g)$. Here $n_g$ is the number of gates
per one map iteration and for polynomial algorithms
$n_g \sim n_q^\gamma$ 
({\it e.g.} for the quantum saw-tooth map 
$\gamma=2$ \cite{benentist}).

Contrary to the case of random errors the effects of static
imperfections on fidelity decay have been studied only in 
\cite{benentist,wavelet}. The numerical simulations performed 
there with up to 18 qubits show that for small strength of static
imperfections $\varepsilon$ the time scale $t_f$ varies as 
$t_f \sim 1/(\varepsilon n_g  \sqrt{n_q})$. Such a dependence
implies that
in the limit of small $\varepsilon$ the effects of static imperfections
dominate the fidelity decay comparing to the case of random errors 
\cite{benentist,wavelet}. Simple estimates based on the Rabi oscillations
have been proposed to explain the above dependence extracted
from numerical data \cite{benentist,wavelet}. 

Since the numerical results show that the static imperfections
lead to a more rapid fidelity decay, compared to random errors
fluctuating from gate to gate,
it is important to investigate their effects in more detail.
 This is the  aim of this paper in which we carry out extensive
numerical and analytical studies of  static imperfections effects on
fidelity decay using as an example a quantum algorithm for
the quantum tent map which describes dynamics in a mixed phase space
with chaotic and integrable motion. For the case when the algorithm
describes the dynamics in the regime of quantum chaos
a scaling theory for universal fidelity decay is developed on the basis 
of the random matrix theory (RMT) \cite{dyson,mehta,guhr}. This theory
is tested in extensive numerical simulations with up to 18 qubits
and the obtained results confirm its analytical predictions
which are rather different from the conclusions of the
previous studies \cite{benentist,wavelet}. We also investigate
the regime of fidelity decay for integrable quantum dynamics
where the situation happens to be more complicated.
In addition, a simple quantum algorithm is proposed
for approximate computation of the coarse-grained Wigner function
(the Husimi function) \cite{wigner,husimi} 
and its stability in respect to imperfections is tested 
on the example of quantum tent map.

It is important to note that all quantum operations required for
implementation of the quantum tent map
have been already realized for 3 - 7 qubits 
in the NMR-based quantum computations reported in Refs. 
\cite{chuang1,cory}. An efficient measurement procedure for fidelity 
decay in quantum computations is proposed in \cite{emerson}.

The paper is organized as follows. In Sec.~2 we describe the
classical and quantum tent map. The algorithm for quantum dynamics is
derived in Sec.~3. The fidelity decay for random errors in quantum gates
is analyzed in Sec.~4. The analytical theory for fidelity decay
induced by static imperfections is developed on the basis of  RMT
approach in Sec.~5. This theory is tested in extensive numerical
simulations presented in Sec.~6. An approximate algorithm for the
quantum computation of the Husimi function is studied in Sec.~7.
The conclusion is given in Sec.~8.  

\vspace{0.5cm}

\section{Classical and quantum tent map}

\label{sec2}

We consider a kicked rotator whose dynamics is governed by the 
time dependent Hamiltonian,
\begin{equation}
\label{eq1}
H(t)=\frac{T p^2}{2}+V(\theta)\sum_{m=-\infty}^\infty \delta(t-m)
\end{equation}
with the potential of kick
\begin{equation}
\label{eq2}
V(\theta)=
\left\{\begin{array}{ll}
-\frac{k}{2}\theta(\theta-\pi) &,\quad 0\le \theta<\pi  \\
\frac{k}{2}(\theta-\pi)(\theta-2\pi) &,\quad \pi\le \theta<2\pi  \\
\end{array}\right.
\end{equation}
where $\theta$ is taken modulo $2\pi$ and $\delta(t)$ 
is a $\delta$-function, $m$ is an integer. 
The parameter $k$ determines the kick strength  
and $T$ gives the rotation of phase between kicks. 
It is easy to see that the classical evolution 
for a finite time step $t \to t+1$ with respect to the 
Hamiltonian (\ref{eq1}) can be described by the map,
\begin{equation}
\label{eq3}
\bar p=p-V'(\theta)\quad,\quad \bar\theta=\theta+\bar p\,T \;
(\texttt{\ mod\ } 2\pi)\quad.
\end{equation}
Here bars mark new values of the dynamical variables after one
map iteration.
This map is similar in structure to the Chirikov standard map 
\cite{chirikov79}. 
The derivative of the kick-potential, 
\begin{equation}
\label{eq4}
V'(\theta)=
\left\{\begin{array}{ll}
k(\frac{\pi}{2}-\theta) &,\quad 0\le \theta<\pi  \\
k(-\frac{3\pi}{2}+\theta) &,\quad \pi\le \theta<2\pi  \\
\end{array}\right.\ ,
\end{equation}
has a tent form and 
is continuous but not differentiable at $\theta=0$ and $\theta=\pi$. 
This is an intermediate case between the standard map \cite{chirikov79} 
with a perfectly smooth kick-potential and the saw-tooth map 
\cite{benentist} with a non-continuous potential. 

\begin{figure}[h]
\begin{center}
\includegraphics[width=0.48\textwidth]{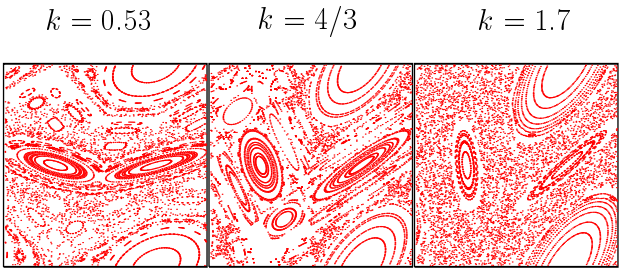}
\end{center}
\caption{\label{fig1} 
Classical Poincar\'e sections of the map (\ref{eq3}) in 
($\theta$, $p$) plane 
for $T=1$ and $K=k=0.53$, $K=k=4/3$ and $K=k=1.7$.
}
\end{figure}

The dynamics of the classical tent map (\ref{eq3})
depends only on one dimensionless
parameter $K=kT$, its properties have 
been studied in \cite{tent,vecheslavov}. 
For small values of $K$ the dynamics is governed 
by a KAM-scenario with the Kolmogorov-Arnold-Moser (KAM) 
invariant curves and a stable island at $\theta=3\pi/2$, $p=0$ and a chaotic 
layer around separatrix starting from  
the unstable  fixed point (saddle) 
at $\theta=\pi/2$, $p=0$. At $K=4/3$, the last 
invariant curve is destroyed and one observes a transition to global chaos 
with a mixed phase space containing big regions with regular dynamics
\cite{tent,vecheslavov}. 

In Fig. \ref{fig1}, the Poincar\'e sections of the map (\ref{eq3}) 
for the three values $K=0.53,\ 4/3,\ 1.7$ are shown.
Here we have replaced $p$ by its value modulo $2\pi/T$ which is 
appropriate since the classical map is invariant with respect to the shift 
$p\to p+2\pi/T$. Fig. \ref{fig1} confirms the above  scenario 
of a transition to global chaos at $K=4/3$. In the following, we are 
particularly interested in a typical  case $K=1.7$, which exhibits global chaos  
with 
quite large stable islands in phase space related to 
the main and secondary resonances. 

At $K \ge 4$, the phase space becomes completely chaotic and the dynamics 
is characterized by a diffusive growth in $p$. The diffusion rate $D$ in
$p$ can be  obtained with the help of random phase approximation
that gives  
\begin{equation}
\label{eq5}
\langle(p-p_0)^2\rangle \approx D\,t\quad,\quad
D=\langle V'(\theta)^2\rangle_\theta=\frac{\pi^2}{12}\,k^2\ .
\end{equation}
Here and below $t$ is an integer which gives the number of map iterations
(kicks). 

\begin{figure}[h]
\begin{center}
\includegraphics[width=0.48\textwidth]{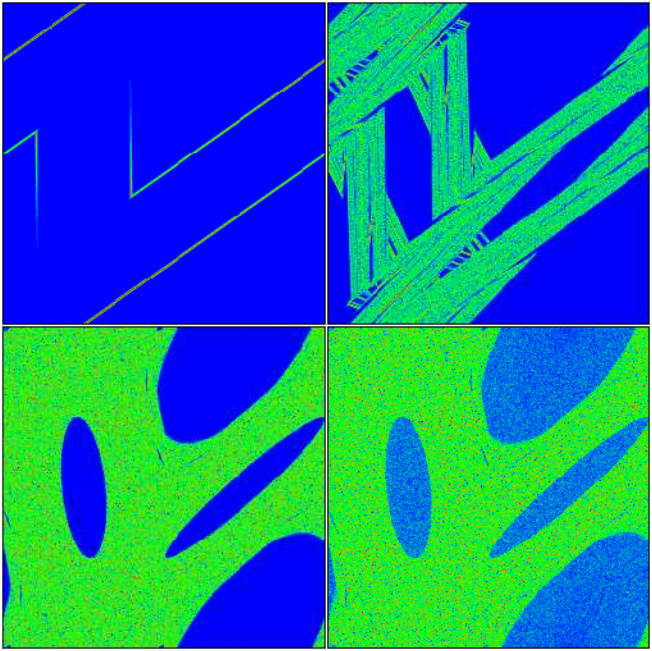}
\end{center}
\caption{\label{fig2} 
Density plots of the Husimi functions 
(see below Sec.~7) of the quantum state 
$|\psi(t)\>$ for $t=5$ (top left panel), $t=15$ (top right panel) and 
$t=5625$ (bottom left panel) with an initial state $|\psi(0)\>$ 
chosen as a minimal coherent (gaussian) wave packet   
closely located to the saddle at $\theta=\pi/2$, $p=0$. 
The bottom right panel corresponds to the quantum state 
after 5625 iterations computed on a quantum computer with static 
imperfections for 
$\varepsilon=7\cdot 10^{-7}$, here the fidelity is 
$f=0.9388$ (see below Secs.~3,6 for details). 
The density is minimal for blue/black
and maximal for red/white. 
All panels correspond to $n_q=16$ qubits, i.e. to a finite dimensional 
Hilbert space of dimension $N=2^{16}$, $T=2\pi/N$. 
}
\end{figure}

The quantum dynamics of the Hamiltonian (\ref{eq1}) 
is given by the Schr\"odinger equation,
\begin{equation}
\label{eq6}
i\frac{\partial}{\partial t}|\psi(t)\>=H(t)\,|\psi(t)\>\ .
\end{equation}
Here in the Hamiltonian $H(t)$ the variables 
$\hat p$ and $\hat \theta$ are operators with the commutator
$[\hat p,\hat\theta]=-i$. 
They have integer eigenvalues $p$ for $\hat p$ and real 
eigenvalues $\theta$ in the interval $[0,2\pi[$ for 
$\hat\theta$. As in the classical case one can 
determine the evolution for one map iteration:
\begin{equation}
\label{eq7}
|\psi(t+1)\>=U\,|\psi(t)\>
=e^{-iT\hat p^2/2}\,e^{-iV(\hat \theta)}\,|\psi(t)\> \; .
\end{equation}
Eq. (\ref{eq7}) corresponds to the 
quantized version of the classical map (\ref{eq3}) and defines a 
quantum map that can be efficiently simulated on a quantum computer.
Here $\hbar=1$ and the quasiclassical limit correspond to $T \to 0$,
$k \to \infty$ with $K=kT=const$. 

For quantum dynamics we concentrate 
our studies on the case $K=kT=1.7$ and $T=2\pi/N$ that corresponds to  
evolution on one classical cell (see Fig.~\ref{fig1}) with $N$ quantum 
states. 
As an initial state we use a minimal coherent wave packet
corresponding to a given $N$ which is placed in a chaotic or integrable
component
(near the unstable or stable fixed point at $p=0, \theta=\pi/2$
or $\theta=3\pi/2$).
An example of the Husimi function (see Sec.~7)
for a chaotic case is shown in 
Fig.~\ref{fig2} for different moments.  

\vspace{0.5cm}

\section{Quantum algorithm}

\label{sec3}

The quantum map (\ref{eq7}) can be simulated on a quantum 
computer in a polynomial number of gates. The quantum algorithm
has similarities with those described in \cite{georgeotkr,benentist}.
To perform one iteration of the quantum map (\ref{eq7})
 we represent the states $|\psi\>$ by a quantum register 
with $n_q$ qubits. In particular, the eigenstates $|p\>$ of the momentum 
operator are identified with the quantum-register states 
\begin{equation}
\label{eq8}
|\alpha_0,\alpha_1,\ldots,\alpha_{n_q-1}\>
=|\alpha_0\>_0\,|\alpha_1\>_1\,\ldots\,|\alpha_{n_q-1}\>_{n_q-1}
\end{equation}
where $\alpha_j = 0$ or $1$ and 
\begin{equation}
\label{eq9}
p=\sum_{j=0 }^{n_q-1} \alpha_j\,2^j\ .
\end{equation}
The states $|0\>_j$ and $|1\>_j$ correspond to the two basis states 
of the $j-$th qubit. Obviously, this representation introduces
a Hilbert space of finite dimension $N=2^{n_q}$; 
the operator $\hat p$ has the eigenvalues: $p=0,\ldots, N-1$. 

A quantum computer is a machine that is able to prepare a quantum register  
with a well defined initial condition 
and to perform certain well controlled unitary operations on this 
quantum register. These particular operations are called quantum gates and 
one typically assumes that the quantum computer can be constructed 
with quantum gates that manipulate at most two qubits. Here we use
as elementary gates 
 the phase-shift gates $B_j^{(1)}$ and controlled phase-shift gates 
$B_{jk}^{(2)}$:
\begin{eqnarray}
\label{eq10}
B_j^{(1)}(\phi)\,|p\>&=&e^{i\alpha_j\phi}\,|p\>\quad,\\
\label{eq11}
B_{jk}^{(2)}(\phi)\,|p\>&=&e^{i\alpha_j\alpha_k\phi}\,|p\>\quad,
\quad j\neq k\quad
\end{eqnarray}
where $p$ is of the form (\ref{eq9}). These gates provide a phase factor 
$e^{i\phi}$ if $\alpha_j=1$ for the simple phase-shift or if 
$\alpha_j=\alpha_k=1$ for the controlled phase-shift. 
Using Eq. (\ref{eq9}), one easily verifies 
that the momentum dependent factor of the unitary operator $U$ 
in (\ref{eq7}) can be expressed in terms of these gates by:
\begin{equation}
\label{eq12}
e^{-iT\hat p^2/2}=
\prod_{j<k}^{n_q-1} B_{jk}^{(2)}(-T\,2^{j+k})
\prod_{j=0}^{n_q-1} B_j^{(1)}(-T\,2^{2j-1})\ .
\end{equation}
The situation is different for the phase factor containing the 
kick-potential since this factor is not diagonal in momentum 
representation. It is therefore necessary to transform to the 
basis of eigenstates of the operator $\hat\theta$. For this, 
following \cite{josza} we 
consider the unitary operator $U_{\rm QFT}$ defined by:
\begin{equation}
\label{eq13}
U_{\rm QFT}\,|p\>=\frac{1}{\sqrt{N}}\sum_{\tilde p=0}^{N-1} 
e^{2\pi i\,p\,\tilde p/N}\,|\tilde p\>\ .
\end{equation}
Then the eigenstates of $\hat\theta$ with eigenvalues 
$\theta=\frac{2\pi p}{N}$ are simply given by: $U_{\rm QFT}^{-1}\,|p\>$ 
and more generally the operators $\hat\theta$ and $\hat p$ are related by:
\begin{equation}
\label{eq14}
\hat\theta=U_{\rm QFT}^{-1}\ \left(\frac{2\pi\,\hat p}{N}\right)
\ U_{\rm QFT}\ .
\end{equation}
Using Eqs. (\ref{eq2}), (\ref{eq9}) and (\ref{eq14}) it is straight-forward 
to show that the unitary factor of $U$ containing the kick-potential 
can be written as:
\begin{eqnarray}
\nonumber
e^{-iV(\hat\theta)}&=& U_{\rm QFT}^{-1}
\prod_{j<k}^{n_q-2}\Bigl\{B_{j,k,n_q-1}^{(3)}(-k\pi^2\,2^{j+k-2n_q+3})\\
\label{eq15}
&& \times \quad B_{jk}^{(2)}(k\pi^2\,2^{j+k-2n_q+2})\Bigr\}\\
\nonumber
&& \times \prod_{j=0}^{n_q-2}\Bigl\{B_{j,n_q-1}^{(2)}(-k\pi^2\,
2^{j-n_q+2}(2^{j-n_q+1}-1))\\
\nonumber
&& \times \quad B_j^{(1)}(k\pi^2\,2^{j-n_q}(2^{j-n_q+1}-1))\Bigr\}
\ U_{\rm QFT}\ .
\end{eqnarray}
Here we have used a three-qubit gate for a controlled-controlled phase-shift 
defined by (cf. Eqs. (\ref{eq10}), (\ref{eq11})):
\begin{equation}
\label{eq16}
B_{jkl}^{(3)}(\phi)\,|p\>=e^{i\alpha_j\alpha_k\alpha_l\phi}\,|p\>\quad.
\end{equation}
In Eq.(\ref{eq15}) it appears for $l=n_q-1$ because of the two distinct 
cases in Eq.(\ref{eq2}). In principle, this gate is not directly 
available in the set of one- and two-qubit gates used for a quantum computer. 
However, it can be constructed from 5 two-qubit gates by:
\begin{equation}
\label{eq17}
B_{jkl}^{(3)}(\phi)=B_{jl}^{(2)}\left(\frac\phi 2\right)
\,B_{jk}^{(2)}\left(\frac\phi 2\right)
\,C^{(N)}_{kl}\,B_{jk}^{(2)}\left(-\frac\phi 2\right)
\,C^{(N)}_{kl}
\end{equation}	
where $C^{(N)}_{kl}$ is the controlled-not gate that exchanges the states 
$|0>_k$ and $|1>_k$ for the $k$-th qubit if $\alpha_l=1$. 
In matrix representation it is given by:
\begin{equation}
\label{eq18}
C^{(N)}_{kl}=\left(\begin{array}{cccc}
1 & 0 & 0 & 0 \\
0 & 1 & 0 & 0 \\
0 & 0 & 0 & 1 \\
0 & 0 & 1 & 0 \\
\end{array}\right)
\end{equation}
where the index $\alpha_l$ corresponds to the outer block structure 
and $\alpha_k$ to the inner block structure. 

Following the description \cite{josza}, the QFT operator $U_{\rm QFT}$ 
can be written in the form:
\begin{equation}
\label{eq19}
U_{\rm QFT}^{\pm 1} = R\ \prod_{j=0}^{n_q-1}\Bigl\{A_j
\prod_{k=j+1}^{n_q-1}B_{jk}^{(2)}(\pm\pi\,2^{j-k})\Bigr\}
\end{equation}
where $R$ is the unitary operator that reverses the order of the qubits 
and $A_j$ is a one-qubit gate with the matrix representation:
\begin{equation}
\label{eq20}
A_j=\frac{1}{\sqrt{2}}
\left(\begin{array}{cc}
1 & 1 \\
1 & -1\\
\end{array}\right)\ .
\end{equation}
We note that in the outer product in Eq.(\ref{eq19}) the factors 
are ordered from left to right with increasing $j$. 

Combining, Eqs.(\ref{eq12}), (\ref{eq15}) and (\ref{eq19}), we see 
that the quantum map (\ref{eq7}) can be expressed by a total number 
of $n_g=\frac92 n_q^2-\frac{11}2 n_q+4$ elementary quantum gates 
(and 2 $R$-operations). On a classical computer one iteration of
the quantum tent map requires $O(n_q 2^{n_q})$ operations
coming mainly from the FFT. 

To investigate the stability of the quantum algorithm for the tent map
we consider two models of imperfections. The first model represents the random
errors in quantum gates fluctuating in time from one gate to another
(random noise errors). In this case for 
all phase-shift gates we replace $\phi$ by $\phi+\delta\phi$ 
with random $\delta\phi\in[-\varepsilon,\,\varepsilon]$ that is 
different for each application of the gate. For the gates containing 
the Pauli matrix $\sigma_x$ we replace it by $\vec n\cdot \vec \sigma$ 
where $\vec n$ is a random unit-vector close to $\vec e_x$ with 
$|\vec n-\vec e_x|\le \varepsilon$. 

The second model describes only static imperfections
and is similar to one used in 
\cite{georgeot,benentist,wavelet,pomeransky}.
In this case the effect of static imperfections is modeled 
by an additional unitary rotation between two arbitrary gates 
which has the form~: $U_s=e^{i \delta H}$. Here the Hamiltonian $\delta H$
is given by: 
\begin{equation}
\label{eq20a}
\delta H=\sum_{j=0}^{n_q-1} \delta_j\, \sigma_j^{(z)}+
2\sum_{j=0}^{n_q-2} J_j\, \sigma_j^{(x)}\,\sigma_{j+1}^{(x)}
\end{equation}
where $\sigma_j^{(\nu)}$ are the Pauli matrices acting on the $j$th qubit 
and $\delta_j,\,J_j\in[-\sqrt{3}\varepsilon,\,\sqrt{3}\varepsilon]$ 
are random coefficients which are drawn only once at the beginning and 
kept fixed during the simulation. These coefficients determine one disorder
realization. In addition to a linear chain of qubits we also 
analyzed a case with qubits distributed on a square lattice
which gave qualitatively similar results (see Sec.~6).

\vspace{0.5cm}

\section{Quantum computation with random errors}
\label{sec4}

\begin{figure}[h]
\begin{center}
\includegraphics[width=0.48\textwidth]{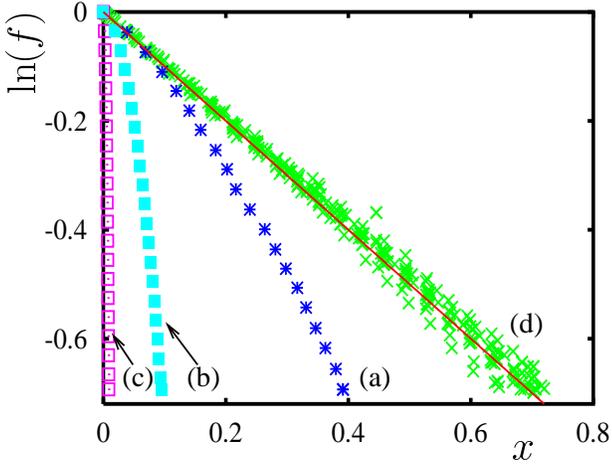}
\end{center}
\caption{\label{fig3} 
Data points (d) represent the 
fidelity decay
for the case of random noise errors in the gates of quantum algorithm.
The fidelity $f=|\<\psi(t)|\psi_{\rm noise}(t)\>|^2$ is shown  
as a function of the scaling variable $x=0.095\,t\,\varepsilon^2\,n_q^2$ 
with $n_q=10, 12, 14, 16$ and $0.001\le \varepsilon\le 0.1$. 
The full line corresponds to the function $f=\exp(-x)$. 
Data points (a), (b) and (c) represent the fidelity decay 
($f=|\<\psi(t)|\psi_{\rm stat}(t)\>|^2$)  
as a function of the scaling variable $x=t\,\varepsilon^2\,n_q n_g^2$ 
for the case of static imperfections with 
$n_q=10$, $n_g=\frac92 n_q^2-\frac{11}2 n_q+4$ being the number of 
elementary gates for one application of the quantum map and 
with (a) $\varepsilon=3\cdot 10^{-5}$, (b) $\varepsilon=6\cdot 10^{-6}$ 
and (c) $\varepsilon=5\cdot 10^{-7}$. The scaling variable $x$ corresponds 
in all cases to $\Gamma t$ where $\Gamma$ is the decay rate obtained 
from Fermi's golden rule. Therefore, the data curves (a), (b), (c) 
are tangent to $\exp(-x)$ for  small $x$. 
The number of shown data points has been strongly reduced in order 
to increase the visibility of the different symbols.
(same approach is used for other figures).
Here and in Figs.~4-8, the initial state is as in Fig.~2, $K=1.7$,
$T=2\pi/N$.
}
\end{figure}

The numerical results for fidelity
decay induced by random errors in quantum gates are presented in 
Fig.~{\ref{fig3}}.  They clearly show that the decrease is exponential
with time $t$ and is given by the fit:
\begin{equation}
\nonumber
f(t) = \exp(-t/t_r) \; ; \
\label{eq4.1}
t_r=1/(0.095 \varepsilon^2 n_q^2) \approx 47/(\varepsilon^2 n_g) \; .
\end{equation}	
As discussed in the introduction, the decay rate per gate
is proportional to  $\varepsilon^2$ since on each step
noise transfer such a probability from an ideal state to all other states
(see \cite{paz,cat,benentist,well,wavelet,bettelli}). 
With a few percent accuracy the numerical
constant in (\ref{eq4.1}) is close to the lower bound discussed 
in \cite{bettelli}. 

While with random errors the fidelity drops by a significant
amount in a purely exponential way the situation
in the case of static imperfections is more complicated (see Fig.~\ref{fig3}).
In this case the initial exponential decrease is followed by
a gaussian exponential one. 
As a result static imperfections give a faster
decay of fidelity. The transition between these two types
of decay depends on the strength of imperfections $\varepsilon$.
Moreover, in scaled variables  the weaker is $\varepsilon$ the stronger is 
the gaussian decrease of fidelity (see Fig~\ref{fig3}).
We shall describe these phenomenon using the following RMT approach.

\section{Quantum computation with static imperfections: RMT approach}
\label{sec5}

Let us denote by $\tilde U$ the unitary operator for the quantum map 
with static imperfections and by $U$ the unitary operator for the ideal 
quantum map. We  denote by $U_j$, $j=1,\ldots, n_g$ the 
elementary quantum gates which constitute the quantum map. According 
to the description of the quantum algorithm in section \ref{sec3} we  
write~:
\begin{equation}
\label{rmt_eq1}
U=U_{n_g}\cdot \ldots\cdot U_2\cdot U_1
\end{equation}
and
\begin{equation}
\label{rmt_eq2}
\tilde U=U_{n_g}\cdot e^{i\delta H}\cdot \ldots\cdot 
U_2\cdot e^{i\delta H}\cdot U_1\cdot e^{i\delta H}
\end{equation}
where $\delta H$ is the hermitian operator describing the static 
imperfections. 
In numerical simulations we have used the particular expression  
(\ref{eq20a}) for $\delta H$ but we mention that our approach 
does not rely on this expression and is much more general. We now introduce 
an effective perturbation operator for the full quantum map by: 
\def\X{{\delta H_{\rm eff}}}
$\tilde U=U\,e^{i\X}$. The operator $\X$ is determined by:
\begin{equation}
\label{rmt_eq3}
e^{i\X}=e^{i\delta H(n_g-1)}\cdot \ldots \cdot e^{i\delta H(1)}
\cdot e^{i\delta H}
\end{equation}
with 
\begin{equation}
\label{rmt_eq4}
\delta H(j)=U_{j-1}^{-1}\cdot\ldots\cdot U_1^{-1}\ \delta H
\ U_1\cdot\ldots\cdot U_{j-1}\ .
\end{equation}
We mention that the precise relation between $\X$ and $\delta H$ is 
not important for the following argumentation and we will need only one 
characteristic time scale $t_c$ defined by:
\def\tr{{\texttt{tr}}}
\begin{equation}
\label{rmt_eq5}
\frac{1}{t_c}=\frac{1}{N}\tr\left(\X^2\right).
\end{equation}
 
We furthermore assume that $\tr(\X)=0$ (the case of $\tr(\X)\neq 0$ 
can be trivially transformed to the this case \cite{prosen} and $\delta H$ 
in Eq. (\ref{eq20a}) has actually a vanishing trace). 

Following Ref. \cite{prosen}, we express the fidelity in terms of 
a correlation function of the perturbation. For this we write 
the fidelity at time $t$ as $f(t)=|A(t)|^2$ 
with the amplitude:
\begin{eqnarray}
\label{rmt_eq6}
A(t)&=&\left\langle U^{-t}\left(U e^{i\X}\right)^t\right\rangle_Q\\
\nonumber
&=&\left\langle e^{i\X(t-1)}\cdot\ldots\cdot e^{i\X(1)}\cdot e^{i\X(0)}
 \right\rangle_Q
\end{eqnarray}
and
\begin{equation}
\label{rmt_eq7}
\X(\tau)=U^{-\tau}\ \X\ U^{\tau}\ .
\end{equation}
Here $\langle\cdots\rangle_Q$ denotes the quantum expectation value. 
For a fixed initial state $|\psi_0\>$ this expectation value is given by: 
$\langle\cdots\rangle_Q=\<\psi_0|\cdots|\psi_0\>$. However, in the following 
we average over all possible initial states that corresponds to:
$\langle\cdots\rangle_Q=\frac{1}{N}\tr(\dots)$. Since we are 
interested in the case where the fidelity is close to 1 
we can expand (\ref{rmt_eq6}) up to second order in $\X$ (or equivalently in 
$\varepsilon$)~:
\begin{eqnarray}
\nonumber
A(t)&\approx& 1+i\sum_{\tau=0}^{t-1}\left\langle\X(\tau)\right\rangle_Q
-\frac{1}{2}\sum_{\tau=0}^{t-1}\left\langle\X^2(\tau)\right\rangle_Q\\
\label{rmt_eq8}
&&-\frac{1}{2}\sum_{\tau_1=0}^{t-1} \sum_{\tau_2=0}^{\tau_1-1} 
\left\langle\X(\tau_1)\X(\tau_2)\right\rangle_Q\ .
\end{eqnarray}
Now we introduce the correlation function $C(\tau)$ by:
\begin{eqnarray}
\label{rmt_eq9}
C(\tau_1-\tau_2)&=&t_c\left\langle\X(\tau_1)\X(\tau_2)\right\rangle_Q\\
\nonumber
&=&t_c\left\langle\X(\tau_1-\tau_2)\X(0)\right\rangle_Q
\end{eqnarray}
and we note that $\left\langle\X(\tau)\right\rangle_Q=\frac{1}{N}\tr(\X)=0$. 
Combining Eqs. (\ref{rmt_eq5}), (\ref{rmt_eq8}) and (\ref{rmt_eq9}), 
we obtain:
\begin{equation}
\label{rmt_eq10}
f(t)\approx
1-\frac{t}{t_c}-\frac{2}{t_c}\sum_{\tau=1}^{t-1} (t-\tau)\,C(\tau)\ .
\end{equation}
This provides the general expression relating the fidelity and the correlation 
function (\ref{rmt_eq9}) previously obtained in Ref. \cite{prosen}. 

We now assume that the unitary quantum map $U$ can be modeled by a random 
matrix drawn from Dyson's  circular orthogonal ($\beta=1$) or unitary 
($\beta=2$) ensemble \cite{dyson,mehta,guhr}. As we will see it 
is useful to express $U$ in terms of its eigenvectors and eigenphases:
\begin{equation}
\label{rmt_eq11}
U=V\,e^{i\hat \theta}\,V^\dagger\quad,\quad
\theta=
\left(\begin{array}{ccc}
\theta_1 &&  \\
 &\ddots &  \\
 & & \theta_N \\
\end{array}\right)\ .
\end{equation}
The matrix $V$ is either real orthogonal ($\beta=1$) or complex unitary 
($\beta=2$). Inserting Eqs. (\ref{rmt_eq7}), (\ref{rmt_eq11}) in 
(\ref{rmt_eq9}), we obtain:
\begin{equation}
\label{rmt_eq12}
C(\tau)=\frac{t_c}{N}\tr\left(V\,e^{-i\tau\hat\theta}\,V^\dagger\,
\X\,V\,e^{i\tau\hat\theta}\,V^\dagger\,\X\right)\ .
\end{equation}
In the following, we want to evaluate the average of $C(\tau)$ with 
respect to $U$. We first evaluate the average with respect to the matrix 
elements of $V$ which gives for $N\gg 1$~:
\begin{equation}
\label{rmt_eq13}
\langle C(\tau)\rangle_U=\left(\frac{2}{\beta}-1\right)\frac{1}{N}
+\frac{1}{N^2}\left\langle\sum_{j,k=1}^N e^{i\tau(\theta_j-\theta_k)}
\right\rangle_\theta\ .
\end{equation}
Here we have used that $\tr(\X)=0$ and 
we have replaced $\tr(\X^2)=N/t_c$ according to Eq. (\ref{rmt_eq5}). 
We have furthermore neglected corrections of order $1/N^2$ which arise 
from small correlations between matrix elements of $V$ at different positions. 
We note that in (\ref{rmt_eq13}) the first term vanishes for $\beta=2$. 
For $\beta=1$ this term arises from additional contributions 
(in the $V$-average) because the elements of $V$ are real for this case. 
The diagonal contributions with $j=k$ in the second term of (\ref{rmt_eq13}) 
provide the constant $1/N$ (which simplifies the first term). 
The average over the non-diagonal contributions with $j\neq k$ 
can be expressed \cite{mehta,guhr} in terms of a double integral over the 
two-point density for the eigenphases $\theta_j$. Since this two-point 
density is related to the two-point correlation function of the 
random matrix theory we obtain for $\tau\ge 1$~:
\begin{equation}
\label{rmt_eq14}
\langle C(\tau)\rangle_U=\frac{1}{N}\left(\frac{2}{\beta}-b_2\left(
\frac{\tau}{N}\right)\right)
\end{equation}
where 
\begin{equation}
\label{rmt_eq15}
b_2(\tilde \tau)=\int_{-\infty}^\infty ds\ Y_2(s)\,e^{2\pi i\,\tilde \tau\,s}
\end{equation}
is the ``two-level form factor'' defined as the Fourier transform of the 
two-point correlation function $Y_2(s)$. The form factor of the 
Wigner-Dyson ensembles is well known \cite{dyson,mehta,guhr}. 
For the unitary and orthogonal symmetry class it reads (in the 
large $N$ limit):
\begin{eqnarray}
\label{rmt_eq16}
\beta&=&2\ :\quad b(\tilde \tau)=\left\{
\begin{array}{ccc}
1-|\tilde \tau| & \ \textrm{if} & |\tilde \tau|\le 1\,, \\
0  &  \ \textrm{if} & |\tilde \tau|> 1\,,\\
\end{array}\right.\\
\nonumber
\beta&=&1\ :\quad b(\tilde \tau)=\left\{
\begin{array}{ccc}
1-2|\tilde \tau|+|\tilde \tau|\ln(2|\tilde \tau|+1) & \ \textrm{if} & |\tilde \tau|\le 1\,, \\
-1+|\tilde \tau|\ln\left(\frac{2|\tilde \tau|+1}{2|\tilde \tau|-1}\right)  &  \ \textrm{if} & |\tilde \tau|> 1\,.\\
\end{array}\right.\\
\label{rmt_eq17}
\end{eqnarray}
Inserting the average correlation function (\ref{rmt_eq14}) in 
(\ref{rmt_eq10}) and replacing the discrete sum by an integral, we finally 
obtain the following scaling expression for the fidelity:
\begin{equation}
\label{rmt_eq18}
-\langle \ln f(t)\rangle_U\approx \frac{N}{t_c}\chi\left(\frac{t}{N}
\right)
\end{equation}
with
\begin{eqnarray}
\label{rmt_eq19}
\chi(s)&=&s+\frac{2}{\beta}\,s^2+\delta \chi(s)\ ,\\
\label{rmt_eq19b}
\delta \chi(s)&=&-2\int_0^s d\tilde \tau\,(s-\tilde \tau)\,b_2(\tilde \tau)
\end{eqnarray}
where $s=t/N$. 
Using the random matrix expressions (\ref{rmt_eq16}), (\ref{rmt_eq17}), 
we find for $\beta=2$:
\def\abstand{{\phantom{\Big|}}}
\begin{equation}
\label{rmt_eq20}
\delta \chi(s)=\left\{
\begin{array}{ccc}
-s^2+\frac{1}{3} s^3 & \ \textrm{if} & s\le 1\,, \abstand\\
-\frac{2}{3}  &  \ \textrm{if} & s> 1\,,\abstand\\
\end{array}\right.
\end{equation}
and for $\beta=1$:
\begin{equation}
\delta \chi(s)=\left\{
\begin{array}{cc}
\left.\begin{array}{l}
\frac{1}{18}(-3s-24s^2+17s^3) \\
+\frac{1}{12}(1+3s-4s^3)\ln(2s+1)\abstand\\
\end{array}\right\} & \textrm{if} \  s\le 1\,, \\
&\\
\left.\begin{array}{l}
\frac34\ln(3)(s-1)-\frac59\\
+\frac13(2-3s+s^2)\abstand\\
+\frac{1}{12}(1-3s+4s^3)\ln(2s-1) \\
+\frac{1}{12}(1+3s-4s^3)\ln(2s+1) \abstand\\
\end{array}\right\} & \textrm{if} \  s> 1\,.\\
\end{array}\right.\\
\label{rmt_eq21}
\end{equation}
Eqs. (\ref{rmt_eq18}-\ref{rmt_eq21}) provide the key result of this 
section. From the practical point of view the contribution of $\delta\chi(s)$ 
in (\ref{rmt_eq19}) is not very important, since (for $\beta=1$):
\begin{eqnarray}
\label{rmt_eq22}
\delta \chi(s)&\approx& -s^2+\frac{2}{3}s^3\quad(\textrm{if}\quad s\ll 1)\ ,\\
\nonumber
\delta \chi(s)&\approx& \left(\frac34\ln(3)-1\right)\,s+\frac16\ln(2s)-
\frac34\ln(3)+\frac13\\
&\approx& -0.17604 s+\frac16\ln(2s)-0.49063\\
\nonumber
&&\qquad\qquad (\textrm{if}\quad s\gg 1)\ .
\label{rmt_eq23}
\end{eqnarray}
Therefore for small $s$ the linear term and for large $s$ the 
quadratic term dominate the behavior of $\chi(s)$ in the expression 
(\ref{rmt_eq19}). In the next section we compare this theoretical 
random matrix result to the numerical data of the fidelity obtained 
for the quantum version of the tent map. 

Before doing so, we want to discuss three particular points. First, we have 
to evaluate the time scale $t_c$ that characterizes the effective strength 
of the perturbation. From Eqs. (\ref{rmt_eq3}) and (\ref{rmt_eq5}), 
we obtain in lowest order in $\varepsilon$:
\begin{equation}
\label{rmt_eq24}
\frac{1}{t_c}=\frac1N \sum_{j,k=0}^{n_g-1} \tr\Bigl(
\delta H(j)\delta H(k)\Bigr)
\end{equation}
with $\delta H(j)$ given by (\ref{rmt_eq4}). This expression is 
similar in structure to Eqs. (\ref{rmt_eq8}) or (\ref{rmt_eq10}) but 
with the important difference that here the ``time'' index corresponds 
to the number of elementary gates and not to the iteration number 
of the quantum map. Since the elementary gates affect only one or two 
qubits (``spins'') the correlation decay between $\delta H(j)$ 
and $\delta H(k)$ will be quite weak and we can obtain a good estimate of 
(\ref{rmt_eq24}) by:
\begin{equation}
\label{rmt_eq25}
\frac{1}{t_c}=a n_g^2 \frac1N \tr(\delta H^2)\approx 
a n_g^2\, 5\,n_q \varepsilon^2
\end{equation}
where $a<1$ is a numerical constant taking into account the exact 
correlation decay and the trace has been evaluated using Eq. (\ref{eq20a}). 
The numerical results of the next Section indicate clearly that 
$a\approx 1/5$ such that the overall numerical factor is 1 and we have:
\begin{equation}
\label{rmt_eq26}
t_c=\frac{1}{\varepsilon^2 n_q n_g^2}\quad,\quad 
n_g=\frac92 n_q^2-\frac{11}2 n_q+4\ .
\end{equation}

The second point to discuss concerns the fact that the phase space is mixed 
and not completely chaotic. As it can be seen from Fig. \ref{fig2} 
the chaotic region fills approximately a fraction of $0.65$ of the full 
phase space. If the initial state is a gaussian wave packet  
placed in the chaotic region then its penetration
inside the integrable islands induced by quantum tunneling
 will take exponentially  long time 
scale: $\propto \exp(N)$. 
Therefore with a good approximation we may say that in absence of imperfections
the dynamics takes place only 
inside the chaotic component. Let us introduce now  the Heisenberg time scale
$t_{\rm H}=2^{n_q}$ which is determined by an average energy level-spacing
for $2^{n_q}$ quantum levels in the whole phase space.
If the whole phase space is chaotic then in the above RMT approach
$N=t_{\rm H}$. However, for the case of Fig.~\ref{fig2}
the chaotic component covers only a relative fraction $\sigma=0.65$
of the whole phase space. Due to that in the previous RMT expressions
we should put $N \approx \sigma t_{\rm H} = 0.65 \cdot 2^{n_q}$ to determine
properly the number of chaotic states.
As a consequence, the expression (\ref{rmt_eq18}) now reads ($\beta=1$  for the tent map):
\begin{equation}
\label{rmt_eq27}
-\langle \ln f(t)\rangle_U \approx \frac{\sigma\, t_{\rm H}}{t_c}
\chi\left(\frac{t}{\sigma\, t_{\rm H}}\right)
\approx \frac{t}{t_c}+\frac{2}{\sigma}\,\frac{t^2}{t_c t_{\rm H}} \; .
\end{equation}
Here we have neglected the contribution of $\delta\chi(s)$.

It is interesting to note that there is a formal analogy between 
the fidelity decay given by (\ref{rmt_eq27}) and the decrease
of the probability to stay near the origin
which has been studied in mesoscopic and RMT systems 
(see {\it e.g.} \cite{khmelnitski,mirlin,maspero,sokolov,frahm}).
There, the time scale $t_c$ is replaced by the diffusive 
Thouless time scale $t_{\rm Th}$ and the second term with the Heisenberg 
time scale has negative sign. 

Finally, the third point concerns the fact that 
the above results are based on the assumption that the quantum evolution 
given by the exact quantum algorithm can be described by RMT. In particular, 
we assume that in the dynamical evolution the ergodicity is established 
very rapidly after a few map iterations. This is correct for the 
choice $T=2\pi/N$ which corresponds to the dynamics in one classical cell. 
We note that 
it is also possible to have many classical cells by the alternative 
choice $T=2\pi L/N$ with $L\gg 1$ but fixed in the semiclassical limit 
$N\to\infty$. For $K$ above the global chaos border, the classical 
dynamics is governed by a diffusive dynamics which covers all cells 
after the Thouless time scale $t_{\rm Th}\approx N^2/D\sim L^2/K^2$ 
where $D$ is the diffusion constant given by Eq. (\ref{eq5}). 
In this case the theoretical treatment has to be modified since 
the matrix $U$ will not be a member of the circular ensemble. 
However, choosing a static perturbation sufficiently complicated such that 
it can be modeled by a random matrix, one can show that the 
relations (\ref{rmt_eq18})-(\ref{rmt_eq19b}) relating the fidelity 
to the two-level form factor are still valid. The two-point energy level 
correlation function for diffusive metals has been calculated in the 
frame work of diagrammatic perturbation theory by Altshuler and Shklovskii 
\cite{altshuler} (see also the review \cite{mirlin}). The two-level form 
factor is now given by $b_2(\tilde \tau)=b_{2,\rm RM}(\tilde \tau)+b_{2,\rm diff.}(\tilde \tau)$ where 
$b_{2,\rm RM}$ denotes the random matrix expressions (\ref{rmt_eq16}) 
or (\ref{rmt_eq17}) and $b_{2,\rm diff.}(\tilde \tau)$ is the correction due 
to diffusive dynamics which is given for a cubic sample by 
\begin{equation}
\label{diff_eq1}
b_{2,\rm diff.}(\tilde \tau)=-\frac{2}{\beta}\sum_{n_1,\ldots,n_d=0 
\atop n_1^2+\ldots n_d^2>0}^{\infty}
\tilde \tau\,e^{-\pi^2 g(n_1^2+\ldots n_d^2)\,\tilde \tau}\ .
\end{equation}
Here $d$ is the spatial dimension and $g$ the dimensionless 
conductance with $g\sim t_{\rm H}/t_{\rm Th}$ and $t_{\rm Th}$ being 
the diffusive Thouless time. In the limit $\tilde \tau\ll g^{-1}$ 
(corresponding to: $t\ll t_{\rm Th}$ since $\tilde\tau=t/N$ with 
$N\sim t_{\rm H}$) the sum 
can be approximated by an integral:
\begin{equation}
\label{diff_eq2}
b_{2,\rm diff.}(\tilde \tau)\approx -\frac{2}{\beta} \frac{\tilde \tau^{1-d/2}}{(4\pi g)^{d/2}}\ .
\end{equation}
Inserting this in (\ref{rmt_eq19b}), we obtain the following diffusive 
correction to the scaling function (\ref{rmt_eq19}):
\begin{equation}
\label{diff_eq3}
\delta\chi_{\rm diff.}(s)=\frac{4}{\beta} 
\frac{1}{(2-d/2)(3-d/2)}\,\frac{s^{3-d/2}}{(4\pi g)^{d/2}}\ .
\end{equation}
We note that this contribution dominates the nonlinear RMT correction 
to $\chi(s)$ in (\ref{rmt_eq19}) 
for $d=1,2,3$ if $\beta=2$ and for $d=3$ if $\beta=1$. The fidelity 
itself is slightly reduced by the diffusive correction according to~:
\begin{equation}
\label{diff_eq4}
f(t)=f_{\rm RM}(t)\,\exp\left(- B\,
\frac{t^{3-d/2}\, t_{\rm Th}^{d/2}}{t_c\,t_{\rm H}^2}\right)
\end{equation}
where $B$ is a positive numerical constant of order one.
We mention that this interesting signature of the Altshuler-Shklovskii 
corrections for diffusive quantum systems in the fidelity decay is 
in principle accessible to efficient quantum computation. 
For our case with $\beta=1$ and $d=1$ this 
correction is small. However for general quantum algorithms with 
diffusive behavior it may be important. The fact that this correction 
reduces the fidelity agrees with the observation that the reduction 
of the volume of the chaotic component ($\sigma$) also leads to 
faster fidelity decay according to Eq. (\ref{rmt_eq27}).

\section{Quantum computation with static imperfections: numerical results}
\label{sec6}

\def\eps{\varepsilon}

\begin{figure}[h]
\begin{center}
\includegraphics[width=0.48\textwidth]{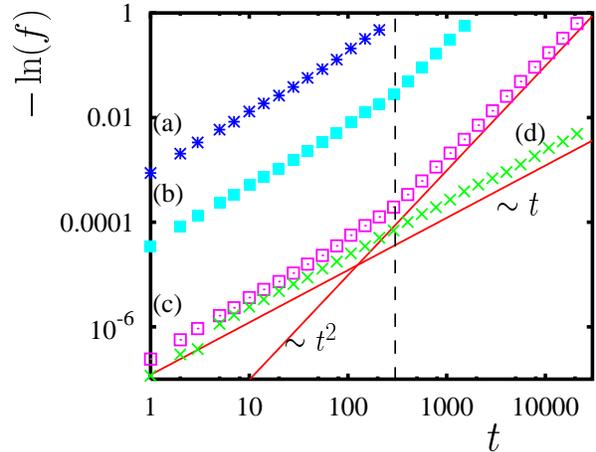}
\end{center}
\caption{\label{fig4} 
The fidelity decay for static imperfections 
[curves (a), (b), (c) with same values for $n_q$ and $\varepsilon$ as in 
Fig. \ref{fig3}] and random errors [curve (d) with $n_q=10$ and 
$\varepsilon=1.59\cdot 10^{-4}$] as a function of $t$ in a double logarithmic 
representation. The full lines correspond to power laws: $-\ln(f)\sim t$ and:
$-\ln(f)\sim t^2$. The value of $\varepsilon$ for (d) is chosen such that 
the average reduction of fidelity for one elementary quantum gate is the 
same as for (c). The vertical dashed line provides the approximate position 
$0.5\sigma t_{\rm H}$
where the curves (a), (b), (c) change from linear to quadratic behavior. 
}
\end{figure}

We now consider the precise model of static imperfections given 
by Eq. (\ref{eq20a}). 
We have numerically calculated the fidelity $f(t)$ for the 
tent map with $K=1.7$ for $n_q\in\{6,\,8,\,10,\,12,\,14,\,16,\,18\}$ and
$5\cdot 10^{-7}\le \eps \le 10^{-4}$. For most cases we have determined 
the fidelity decay up to time scales $t\le t_{max}$ with 
$f(t_{max})=0.5$ (except for $n_q=18$ and the smallest values 
of $\eps$) since we are mostly interested in the 
regime $(1-f)\ll 1$ for which the analytical theory of the previous 
section is valid. We have also considered values $\eps>10^{-4}$ but here 
the value $t_{max}$ is typically so small that the number of available 
data points is not useful for the scaling analysis given below. In most cases 
we have concentrated on one particular realization of the random 
coefficients $\delta_i$ and $J_i$. 
But  we also have made checks with up to 200 particular realizations.
As initial state $|\psi(t=0)\>$ 
we have chosen a coherent state $|\varphi(p_0,\theta_0)\>$ [see 
next section, Eqs. (\ref{eq23}), (\ref{eq24})] which is 
quite well localized around a classical point $(p_0,\theta_0)$ in 
phase space with a relative width $\sim 1/\sqrt{N}\sim 2^{-n_q/2}$ in 
both directions. 

First, we chose a state close to the hyperbolic fix point 
$\theta=\pi/2$, $p=0$, well inside the chaotic region of phase space. 
As can be seen in Fig. \ref{fig2} after $t=15$ iterations the state 
fills up a big fraction of the chaotic region and after $t\approx 30$ 
the state is practically ergodic. It covers then a fraction 
$\sigma\approx 0.65$ of phase space. 

\begin{figure}[h]
\begin{center}
\includegraphics[width=0.48\textwidth]{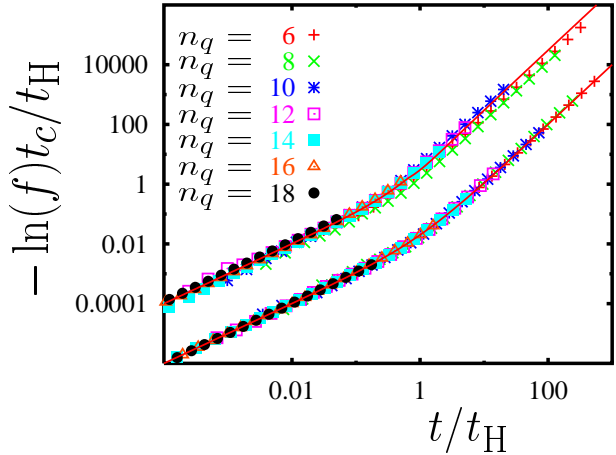}
\end{center}
\caption{\label{fig5}
Scaling representation of the fidelity $f$  for one particular realization of
static imperfections. The upper scaling curve shows: $-\ln(f) t_c/t_{\rm H}$ 
as a function of $t/t_{\rm H}$ with the two theoretical time scales 
$t_c=(\eps^2 n_q n_g^2)^{-1}$ and $t_{\rm H}=2^{n_q}$. 
The full line in the upper curve corresponds to the theoretical 
random matrix result (\ref{rmt_eq18}), (\ref{rmt_eq19}) for $\beta=1$. 
The lower scaling curve (shifted down by a factor 0.01) correspond 
to $-\ln(f) \tilde t_c/\tilde t_{\rm H}$ versus $t/\tilde t_{\rm H}$ 
with the times scales $\tilde t_c$ and $\tilde t_{\rm H}$ 
obtained from the fit (\ref{nfid_eq1}) (using appropriate weight-factors; 
see text) for each value of $n_q$ and $\varepsilon$. 
Here the full line corresponds to the analytical scaling curve: $y=x+x^2$ 
(with $y=-\ln(f) \tilde t_c/\tilde t_{\rm H}$ and $x=t/\tilde t_{\rm H}$). 
Data points are shown 
for $\varepsilon=5\cdot 10^{-7}$ (and $6\le n_q \le 18$);
points for other values of $\varepsilon$ with 
$5\cdot 10^{-7}\le \varepsilon \le 10^{-4}$ fall on the same 
scaling curve and are not shown.
}
\end{figure}

We have already seen in Fig. \ref{fig3} of section \ref{sec4} that the 
fidelity decay for static imperfections is faster 
than the exponential behavior for random errors. In order to 
analyze this in more detail we show in a double logarithmic 
representation in Fig. \ref{fig4}: $-\ln(f(t))$ as 
a function of $t$ for the three cases already shown in Fig. \ref{fig3}  
($n_q=10$ and $\eps=3\cdot 10^{-5},\,6\cdot 10^{-6},\,5\cdot 10^{-7}$). 
For comparison, we also provide one case for random errors 
($n_q=10$ and $\eps=1.59\cdot 10^{-4}$). 

For the static imperfections, we can clearly identify a transition 
from linear to quadratic behavior at a time scale 
$0.5\sigma t_{\rm H}\approx 0.325 t_{\rm H}$ corresponding to 
the theoretical expression (\ref{rmt_eq27}). However, the quadratic 
regime is best visible for the smallest values of $\eps$ due to 
the restriction $-\ln(f)\le \ln(2)$. For the case of random errors 
there is no such transition and the linear behavior applies for all 
time scales. 

To analyze this transition in a quantitative way we determine 
for each value of $n_q$ and $\eps$ two 
time scales $\tilde t_c$ and $\tilde t_{\rm H}$ by the numerical fit 
\begin{equation}
\label{nfid_eq1}
y(t)=\frac{t}{\tilde t_c}+\frac{t^2}{\tilde t_c 
\tilde t_{\rm H}}\quad,\quad y(t)=-\ln(f(t))\ .
\end{equation}
In order to prevent  this fit to be artificially dominated by the 
large values of $t$ (i.e. the quadratic regime) we minimize:
\begin{equation}
\label{nfid_eq2}
d(a_0,a_1)=\sum_t w(t)\left[y(t)-a_0 t-a_1 t^2\right]^2
\end{equation}
with an appropriate weight factor $w(t)\sim 1/(t\,y^2(t))$. The 
factor $1/y^2$ ensures that the vertical distance to be minimized is 
measured in the logarithmic representation for $y$. The other factor 
$1/t$ takes into account that the horizontal density of data points 
in the logarithmic representation increases with $t$. The fit procedure 
(\ref{nfid_eq2}) corresponds therefore to a fit in log-log representation 
such that also the small time scales (and values of $y$) are taken 
properly into account. 

\begin{figure}[h]
\begin{center}
\includegraphics[width=0.48\textwidth]{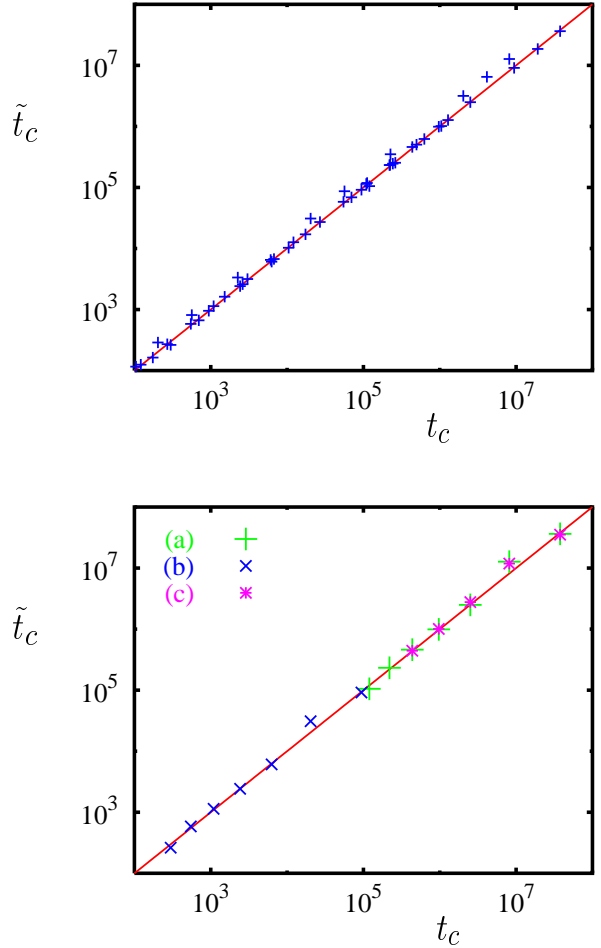}
\end{center}
\caption{\label{fig6} 
The time scale $\tilde t_c$ obtained from the fit: 
$-\ln(f)=t/\tilde t_c+t^2/(\tilde t_c \tilde t_{\rm H})$ versus 
the theoretical expression $t_c=(\varepsilon^2 n_q n_g^2)^{-1}$ in 
a double logarithmic representation. 
The full line corresponds to $\tilde t_c=t_c$. The data points 
of top panel correspond to the same realization of static imperfections 
as in Fig. \ref{fig5} with $6\le n_q \le 18$ and 
$5\cdot 10^{-7}\le \varepsilon \le 10^{-4}$. 
The bottom panel shows again the data points for $6\le n_q \le 18$ and 
$\varepsilon=5\cdot 10^{-7}$ (a) and $\varepsilon=10^{-5}$ (b) 
for the same realization. The data points (c) 
are obtained from $\langle \tilde t_c^{-1}\rangle^{-1}$ 
where $\langle\cdots\rangle$ denotes the average over 200 realizations 
of static imperfections 
for each value of $n_q=6,\,8,\,10,\,12,\,14$ and $\varepsilon=5\cdot 10^{-7}$
(statistical error bars are smaller than symbol size). 
}
\end{figure}

\begin{figure}[h]
\begin{center}
\includegraphics[width=0.48\textwidth]{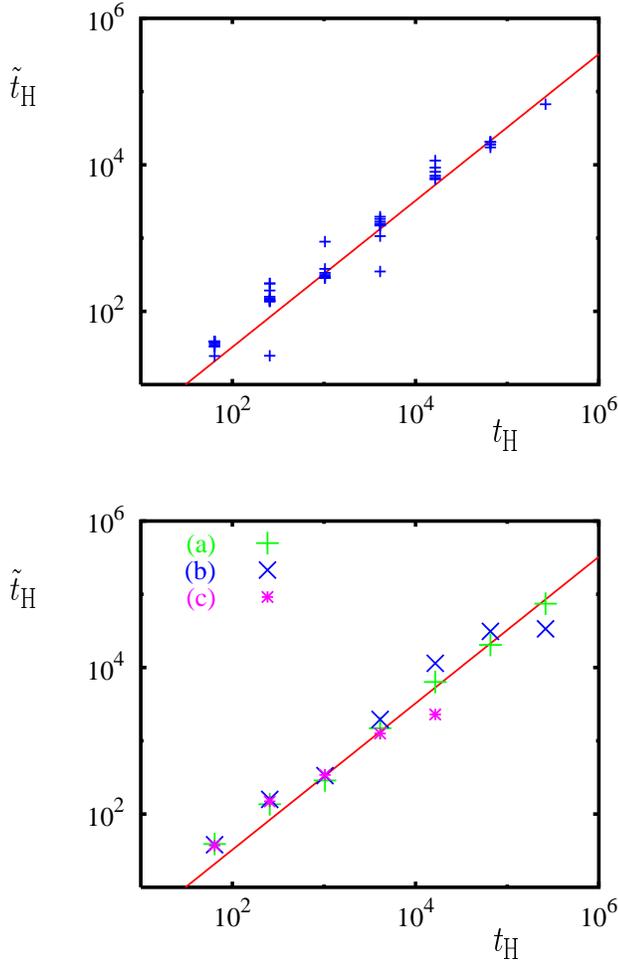}
\end{center}
\caption{\label{fig7} 
The time scale $\tilde t_{\rm H}$ obtained from the fit: 
$-\ln(f)=t/\tilde t_c+t^2/(\tilde t_c \tilde t_{\rm H})$ versus 
the Heisenberg time $t_{\rm H}=2^{n_q}$ in a double logarithmic representation. 
The full line corresponds to $\tilde t_{\rm H}=0.5 \sigma t_{\rm H}$. 
The data points in the top panel 
 correspond to the same realization of static imperfections 
as in Fig. \ref{fig5} with $6\le n_q \le 18$ and 
$5\cdot 10^{-7}\le \varepsilon \le 10^{-4}$. 
The bottom panel shows again the data points for $6\le n_q \le 18$ and 
$\varepsilon=5\cdot 10^{-7}$ [data points (a)] and $\varepsilon=10^{-5}$ 
[data points (b)] for the same realization. The data points (c) 
are obtained from 
$\langle \tilde t_c^{-1}\rangle 
\langle (\tilde t_c \tilde t_{\rm H})^{-1}\rangle^{-1}$
where $\langle\cdots\rangle$ denotes the average over 200 realizations 
of static imperfections 
for each value of $n_q=6,\,8,\,10,\,12,\,14$ and $\varepsilon=5\cdot 10^{-7}$
(statistical error bars are smaller than symbol size). 
}
\end{figure}

\begin{figure}[h]
\begin{center}
\includegraphics[width=0.48\textwidth]{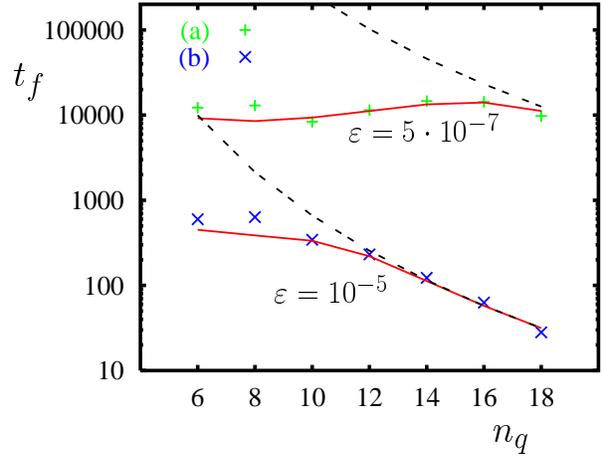}
\end{center}
\caption{\label{fig8} 
The time scale $t_f$ determined by $f(t_f)=0.9$ 
(in a logarithmic representation) as a function of the number of qubits $n_q$. 
The data points correspond to the numerical simulation for the same 
realization of static imperfections as in Fig. \ref{fig5} with 
$\varepsilon=5\cdot 10^{-7}$ (a) and $\varepsilon=10^{-5}$ (b). 
The two full lines correspond to $t_f$ given by (\ref{nfid_eq4}) 
assuming the theoretical expression 
Eq.~(\ref{rmt_eq27}) for the fidelity. 
The two dashed curves correspond to $t_f=t_c\,\ln(\frac{10}{9})$ 
for the simplified exponential 
behavior $f(t)=\exp(-t/t_c)$. 
}
\end{figure}

According to the theoretical expression (\ref{rmt_eq27}) 
one expects that:
\begin{equation}
\label{nfid_eq3}
\tilde t_c=t_c=\frac{1}{\eps^2 n_q n_g^2}\quad,\quad
\tilde t_{\rm H}=0.5\,\sigma\, t_{\rm H}\approx 0.325\,t_{\rm H}
\end{equation}
with $t_{\rm H}=2^{n_q}$ and $n_g=n_q^2-\frac{11}2 n_q+4$. This theoretical 
prediction is verified in Figs.~\ref{fig5}-\ref{fig8}. 

In Fig. \ref{fig5}, we show two types of scaling curves for the fidelity. 
The first (upper) curve shows: $-\ln(f) t_c/t_{\rm H}$ versus $t/t_{\rm H}$ 
with the time scales $t_c$ and $t_{\rm H}$ given above. We observe that 
the numerical data coincide very well for $n_q\ge 10$ with the analytical 
random matrix result (\ref{rmt_eq18}),(\ref{rmt_eq19}) for $\beta=1$. The 
data for $n_q=6,\,8$ show a moderate deviation for $t>t_{\rm H}$. We note 
that for this first scaling curve the dependence of the scaling parameters 
$t_c$ and $t_{\rm H}$ on $\eps$ and $n_q$ is entirely determined by 
their theoretical expressions. This is different for the second (lower) 
scaling curve where: $-\ln(f) \tilde t_c/\tilde t_{\rm H}$ versus 
$t/\tilde t_{\rm H}$ is shown. Here the scaling parameters 
$\tilde t_c$ and $\tilde t_{\rm H}$ have been obtained by the fit 
(\ref{nfid_eq1}) for each value of $n_q$ and $\eps$. Therefore all data 
coincide well with analytical scaling expression (\ref{nfid_eq1}).

It is important to note that both scaling curves cover 10 orders of 
magnitude and provide a strong confirmation of the crossover from linear 
to quadratic behavior predicted by the RMT approach. We mention as 
a side remark that we have also performed a similar scaling analysis for 
the case of random errors. Here the scaling curve is purely linear in 
accordance with Fig. \ref{fig3}. However, this gives a stronger confirmation 
of the linear behavior than in Fig. \ref{fig3} since there the data for small 
$\eps$ and large $n_q$ corresponding to the regime $(1-f)\ll 1$ are
quite badly visible in contrast to the scaling curve.

In Figs. \ref{fig6} and \ref{fig7}, the time scales $\tilde t_c$ and 
$\tilde t_{\rm H}$ obtained from the fit (\ref{nfid_eq1}) are shown versus 
$t_c$ and $t_{\rm H}$. We observe that the first theoretical expectation 
$\tilde t_c=t_c$ is very well verified for the majority of data points. 
The small deviation for the remaining points appear for small $n_q$ and 
the largest values of $\eps$ where the fit procedure is less reliable. 
The second identity $\tilde t_{\rm H}=0.325\,t_{\rm H}$ is in general 
also quite well verified. However, the deviations are slightly larger 
especially for larger values of $\eps$. Furthermore, for $n_q > 14$ the 
regime $t\gg t_{\rm H}$ is numerically not accessible and the fit procedure 
amounts to extrapolate $\tilde t_{\rm H}$ from data points $t\sim t_{\rm H}$ 
or even $t<t_{\rm H}$ for $n_q=16,\,18$. 

We also note that the data points for $n_q=6,\,8$ (lower panel of Fig. 
\ref{fig7}) lie above the theoretical line in accordance with the 
first scaling curve in Fig. \ref{fig5}. 

We have also determined the time scale $t_f$ at which $f(t_f)=0.9$. 
The theoretical expression (\ref{rmt_eq27}) suggests~:
\begin{equation}
\label{nfid_eq4}
t_f=\frac{2 t_c\,\ln(\frac{10}{9})}{1+\sqrt{1+\frac{8}{\sigma}\,
\frac{t_c}{t_{\rm H}}\,\ln(\frac{10}{9})}}
\approx \frac{0.2107\cdot t_c}{1+\sqrt{1+1.2967\,
\frac{t_c}{t_{\rm H}}}}\ .
\end{equation}
For $t_c\ll t_{\rm H}$ this implies $t_f\approx t_c\,\ln(\frac{10}{9})$ 
while for $t_c\gg t_{\rm H}$ we have 
$t_f\sim \sqrt{t_c\,t_{\rm H}}=\sqrt{t_c}\,2^{n_q/2}$. 
In Fig. \ref{fig8}, we show $t_f$ obtained from the numerical data 
for $\eps=10^{-5},\ 5\cdot 10^{-7}$ and all values of $n_q$. 
The data points for $n_q\ge 10$ coincide very well with 
(\ref{nfid_eq4}) while the points for $n_q=6,\,8$ lie slightly 
above the theoretical line. For comparison, we also show the time scale 
$t_f$ obtained from the simplified exponential behavior $f(t)=\exp(-t/t_c)$. 
Generally, $t_f$ is believed to decrease with increasing $n_q$ and fixed 
$\eps$. However, this is not the case if $t_c>t_{\rm H}$, i.e. for 
$\eps<2^{-n_q/2}/(n_g\sqrt{n_q})$. For very small values of $\eps$ 
there is certain regime where $t_f$ first slightly increases with $n_q$ 
and then decreases. 

In all presented numerical studies we 
considered the static couplings between qubits ordered on a line.
To check that the results are not sensitive to this specific configuration we
also considered the case when qubits are located on a square lattice
as it was discussed in \cite{georgeot}. The obtained results (that we do not show here)
confirms the RMT scaling (\ref{rmt_eq27}).

\begin{figure}[h]
\begin{center}
\includegraphics[width=0.48\textwidth]{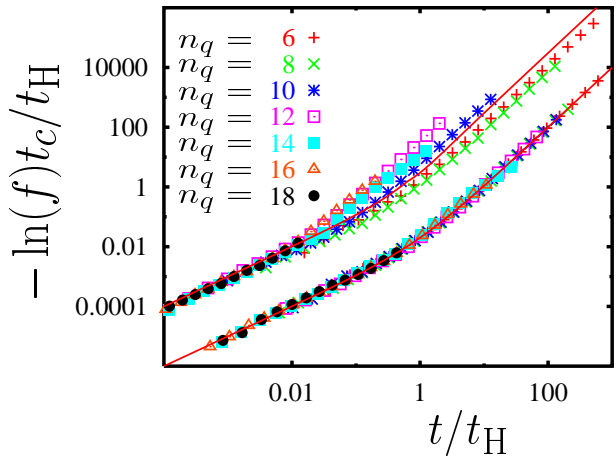}
\end{center}
\caption{\label{fig9} 
Scaling curve for fidelity as in the case
of Fig.~\ref{fig5} but for the 
quantum evolution inside integrable component. Here, the initial state  
is a minimal coherent  wave packet taken 
inside the regular part of phase space
at $\theta=5.35$ and $p=0$; the same realization of static imperfections
as in Fig.~\ref{fig5} is used. 
}
\end{figure}

\begin{figure}[h]
\begin{center}
\includegraphics[width=0.48\textwidth]{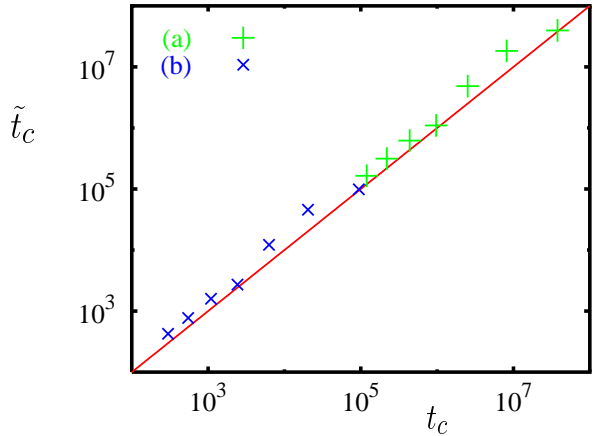}
\end{center}
\caption{\label{fig10} 
The time scale $\tilde t_c$ as in Fig. \ref{fig6} versus 
$t_c=(\varepsilon^2 n_q n_g^2)^{-1}$,
data are obtained from  Fig.~\ref{fig9} for the case of 
 regular dynamics.
Shown are data points for 
$\varepsilon=5\cdot 10^{-7}$ (a) and $\varepsilon=10^{-5}$ (b) 
with $6\le n_q \le 18$ in a double logarithmic representation. 
The full line corresponds to $\tilde t_c=t_c$. 
}
\end{figure}

\begin{figure}[h]
\begin{center}
\includegraphics[width=0.48\textwidth]{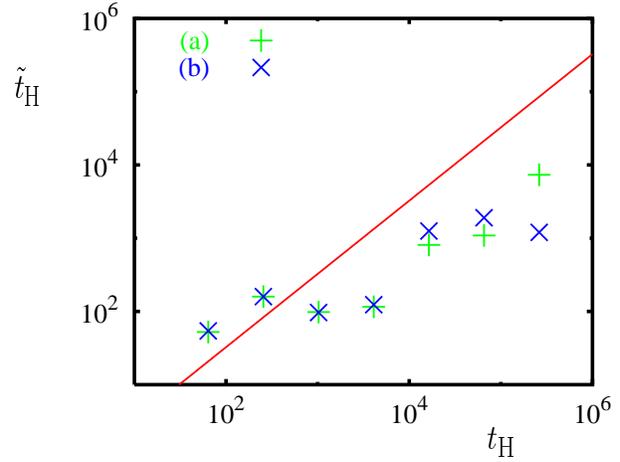}
\end{center}
\caption{\label{fig11} 
The time scale $\tilde t_{\rm H}$ as in Fig. \ref{fig7} versus 
$t_{\rm H}=2^{n_q}$,
data are obtained from  Fig.~\ref{fig9} for the case of 
 regular dynamics.
Shown are data points for 
$\varepsilon=5\cdot 10^{-7}$ (a) and $\varepsilon=10^{-5}$ (b) 
with $6\le n_q \le 18$ in a double logarithmic representation. 
The full line corresponds to $\tilde t_{\rm H}=0.5 \sigma t_{\rm H}$. 
}
\end{figure}

The case of the quantum evolution inside the integrable component of the tent
map is analyzed in Figs.~\ref{fig9}-\ref{fig11}. 
Here, the initial state is located at $\theta=5.35$ and $p=0$ which 
is in middle between the center fix point ($\theta=3\pi/2$, $p=0$) and 
the boundary of the stable island ($\theta\approx 6.0$, $p=0$). 
We have determined for this case the time scales 
$\tilde t_c$ and $\tilde t_{\rm H}$ from the fit (\ref{nfid_eq1}) and 
performed the same scaling analysis in Fig.~\ref{fig9} for 
the fidelity decay as in in Fig.~\ref{fig5} for the initial condition 
in the chaotic component. The scaling curves with the theoretical
expressions (\ref{nfid_eq3}) for $t_c$ and $t_{\rm H}$
give a significant deviation from the RMT result
(upper group of curves in Fig.~\ref{fig9}). To understand the reason
of this dispersion we also show the scaling curves with the 
fitted time scales $\tilde t_c$ and $\tilde t_{\rm H}$.  
This procedure gives a good scaling of numerical data
(lower group of curves in  Fig.~\ref{fig9}).
Obviously, the fit (\ref{nfid_eq1}) still works quite well as such but 
the obtained fit parameters are eventually different from the 
initial condition in the chaotic component and the RMT. 

The dependence of $\tilde t_c$ on the theoretical value of $t_c$
is presented in Fig.~\ref{fig10}. It shows that 
the theoretical expression works with a good accuracy 
in the interval of 6 orders of magnitude. This is not really a surprise since 
according to (\ref{rmt_eq5}) $t_c^{-1}$ measures the overall strength 
of the perturbation. However, for $\tilde t_{\rm H}$
shown in Fig.~\ref{fig11} the situation is much more complicated.
The variation of $\tilde t_{\rm H}$ vs. $t_{\rm H}=2^{n_q}$ shows unusual steps
and it is unclear what is the real dependence in the limit of large $n_q$.
Further studies are required for complete understanding of 
the static imperfection effects in this regime.
This fact compromises the possibility to determine
the asymptotic dependence of $t_f$ on $\varepsilon$ and $n_q$
for the case of integrable or quasi-integrable dynamics.
In addition, the data of Fig.~\ref{fig8} show that 
it is not easy to determine the asymptotic behavior of $t_f$
in absence of clear scaling laws. Due to these two remarks
we think that the scaling dependence for $t_f$ time scale,
proposed in \cite{benentist,wavelet} for the case of static
imperfections, represents in fact only an intermediate 
behavior and cannot be extrapolated to the limit of large $n_q$.
Indeed, the quantum evolutions studied in \cite{benentist,wavelet}
correspond to quasi-integrable regimes and additional tests
are required to check the validity of the RMT scaling
(\ref{rmt_eq27}) for the quantum algorithms studied there.

\begin{figure}[h]
\begin{center}
\includegraphics[width=0.44\textwidth]{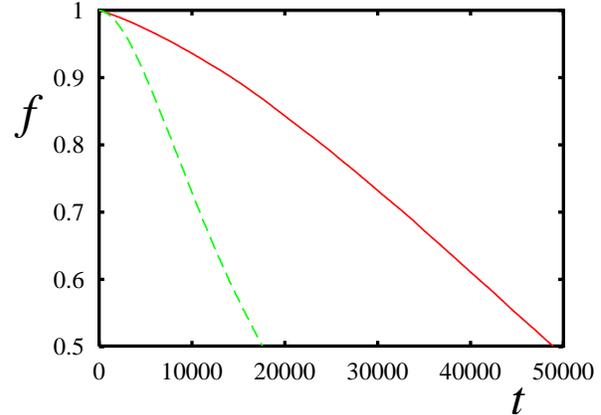}
\end{center}
\caption{\label{fig12} 
fidelity decay for initial minimal coherent state in
chaotic (full curve) and integrable (dashed curve) component
for $n_q=16$, $\varepsilon=5 \cdot 10^{-7}$.
}
\end{figure}

Finally, it is interesting to compare directly the fidelity decay
induced by static imperfections for the quantum evolution
in chaotic and integrable components (see Fig.\ref{fig12}).
The numerical data show that $f(t)$ decreases faster in the case
of integrable evolution. As it was discussed in \cite{prosen}
the presence of chaos reduces the fidelity decay rate.
This is in the agreement with the results of Fig.~\ref{fig7}
and Fig.~\ref{fig11} according to which $\tilde t_{\rm H}$ is much smaller
for the integrable regime as for the regime of quantum chaos.
However, the possibility  of using this fact to improve
the accuracy of quantum computations remains an open question.

The numerical data presented in this section 
definitely confirm the RMT universal law for fidelity decay
for the case when the evolution takes place in the regime of quantum chaos.
This means that this law works for quantum algorithms simulating
a complex dynamics.
The situation for the evolution in the integrable component 
is more complicated. The data show that the theoretical
expression for $t_c$ is still valid but the dependence 
of the time scale $\tilde t_{\rm H}$ on $n_q$  requires further
investigations.

It is interesting to note that the relation (\ref{rmt_eq27}) should also work
for the problem of Loschmidt echo in systems with 
quantum chaos \cite{pastawski,beenakker,como,prosen,cohen}. 
In this case for small perturbations
$t_c$  is still given by Eq.~(\ref{rmt_eq5})
or, that is equivalent, the
inverse decrease rate is given by the Fermi golden rule \cite{beenakker}.
Then the scale $t_{\rm H}$ is determined by the inverse density of states
or for quantum maps by the number of states via relation $t_{\rm H}=N$.
As a result for small perturbations the decay of Loschmidt echo 
for such quantum dynamics is still given by the universal decay
relation Eq.~(\ref{rmt_eq27}).

\section{Husimi function}
\label{sec7}

Here we  discuss how 
an arbitrary quantum state $|\psi\>$ can be represented in the
classical phase space
in the process of quantum computation. 
For this it is convenient to use the coarse-grained
Wigner function (or the Husimi function) \cite{wigner,husimi}:
\begin{equation}
\label{eq22}
\rho_H(p_0,\theta_0)=|\<\varphi(p_0,\theta_0)\,|\,\psi\>|^2
\end{equation}
where the smoothing is done with  the coherent state
\begin{equation}
\label{eq23}
|\varphi(p_0,\theta_0)\>=A\sum_{p} e^{-(p-p_0)^2/4a^2-i\theta_0 p}
\,|p\>\ .
\end{equation}
Here, $A$ is the normalization constant and $a$ is the width of the 
coherent state in the $p$-representation. The coherent state corresponds 
to a gaussian wave packet that is localized in the classical phase space 
around a point $(\theta_0,\,p_0)$ with widths $\Delta p=a$ 
and $\Delta\theta=1/(2a)$. We choose $a=\sqrt{N/12}$ such 
that the widths relative to the size of the phase space are comparable:
\begin{equation}
\label{eq24}
\frac{\Delta p}{N}=\frac{1}{\sqrt{12 N}}\approx\frac{0.2887}{\sqrt{N}}
\quad,\quad \frac{\Delta \theta}{2\pi}=\frac{\sqrt{3}}{2\pi\sqrt{N}}\approx
\frac{0.2757}{\sqrt{N}}\ .
\end{equation}
The naive evaluation of the Husimi function (\ref{eq22}) without any 
optimization 
requires ${\cal O}(N N_p N_\theta)$ operations (on a classical computer) where 
$N_p$ and $N_\theta$ are the numbers of values for $p_0$ and $\theta_0$ 
for which (\ref{eq22}) is evaluated. In view of Eq. (\ref{eq24}) it 
is sufficient to choose $N_p=N_\theta=\sqrt{N}$ resulting in ${\cal O}(N^2)$
operations which is very expensive as compared to ${\cal O}(N\log(N)^2)$ 
operations needed by the simulation of the quantum map 
on a classical computer as described 
in Sec.~\ref{sec3}. 

Fortunately, the evaluation of the Husimi function can be done in a more 
efficient way. To motivate and explain this let us first 
consider a modified Husimi function defined by:
\begin{equation}
\label{eq25}
\rho_{H}^{(p)}(p_0,\theta_0)=|\<\varphi^{(p)}(p_0,\theta_0)\,|\,\psi\>|^2
\end{equation}
with the modified coherent state:
\begin{equation}
\label{eq26}
|\varphi^{(p)}(p_0,\theta_0)\>=\frac{1}{\sqrt[4]{N}}
\sum_{p=p_0}^{p_0+\sqrt{N}-1} e^{-i\theta_0 p}
\,|p\>\ .
\end{equation}
Here we assume for the sake of simplicity that the number of qubits 
$n_q$ is even such that 
$\sqrt{N}=2^{n_q/2}$ is integer. We furthermore require that 
$p_0$ is an integer multiple of $\sqrt{N}$ and 
$\theta_0=2\pi\,l/\sqrt{N}$ with $l=0,\,\ldots,\,\sqrt{N}-1$. 

Comparing (\ref{eq23}) with 
(\ref{eq26}), we see that the gaussian pre-factor has been replaced 
by a box-function of width $\sqrt{N}$. This provides a very good localization 
for the momentum representation but implies that in angle representation the 
amplitude around $\theta_0$ decreases only as a power law 
according to the Fourier transform of the box-function: 
$\sin(x)/x$ with $x=\sqrt{N}(\theta-\theta_0)/2$. However, the modified 
coherent state (\ref{eq26}) still provides a quite well localized state around 
the point $(\theta_0,\,p_0)$. Its main advantage is related to the fact 
that it can be put in the form:
\begin{equation}
\label{eq27}
|\varphi^{(p)}(p_0,\theta_0)\>=\tilde U_{\rm QFT}^{-1}\,|p_0+l\>
\end{equation}
where $\theta_0=2\pi\,l/\sqrt{N}$ and 
$\tilde U_{\rm QFT}$ corresponds to the quantum Fourier transform operator 
(see (\ref{eq13})) 
for the first half of the qubits ($\alpha_0,\,\ldots,\,\alpha_{n_q/2-1}$). 
Eq. (\ref{eq27}) implies that 
\begin{equation}
\label{eq28}
\rho_{H}^{(p)}(p_0,\theta_0)=|\<p\,|\,\tilde U_{\rm QFT}\,|\,\psi\>|^2
\end{equation}
with $p=p_0+l=p_0+\sqrt{N}\theta_0/(2\pi)$. 
Here the state $\tilde U_{\rm QFT}\,|\,\psi\>$ can be evaluated
efficiently on a quantum computer using 
$\frac{n_q}{4}(\frac{n_q}{2}+1)$ elementary quantum gates according to 
(\ref{eq19}) (with $n_q$ replaced by $n_q/2$). Emulating the quantum 
computer on a classical computer this still costs only 
${\cal O}(N\log(N)^2)$ elementary operations. The matrix elements 
$\<p|\,\tilde U_{\rm QFT}\,|\,\psi\>$
of this state with the momentum eigenstates $|p\>$ provide directly 
via Eq. (\ref{eq28}) the modified Husimi function. Here the value of 
$p=0,\,\ldots\,N-1$ contains in its first half of the binary digits the 
information 
for $\theta_0$ and in its second half the information for $p_0$. 
More explicitly, if $p=\sum_{j=0}^{n_q-1} \alpha_j 2^j$, we have~:
\begin{equation}
\label{eq29}
p_0=\sum_{j=n_q/2}^{n_q-1} \alpha_j 2^j \quad,\quad
\theta_0=\frac{2\pi}{\sqrt{N}}\sum_{j=0}^{n_q/2-1}\alpha_j 2^j\ .
\end{equation}

We note that it is also possible to introduce another type of modified 
Husimi function (and modified coherent state) by exchanging the roles of 
$\theta_0$ and $p_0$~:
\begin{equation}
\label{eq30}
\rho_{H}^{(\theta)}(p_0,\theta_0)=|\<p\,|\,\tilde U_{\rm QFT}^{-1}
\,U_{\rm QFT}\,|\,\psi\>|^2\ .
\end{equation}
where $p=N\theta_0/(2\pi)+p_0/\sqrt{N}$. As in Eq. (\ref{eq26}), we require 
that $p_0$ is an integer multiple of $\sqrt{N}$ and 
$\theta_0=2\pi\,l/\sqrt{N}$ with $l=0,\,\ldots,\,\sqrt{N}-1$. 
The operator $U_{\rm QFT}$ corresponds to quantum Fourier transform 
for all qubits and transforms a state from $p$- to $\theta$-representation. 
$\tilde U_{\rm QFT}$ corresponds as above to quantum Fourier transform 
for the first half of the qubits. 
We mention that the coherent states associated to (\ref{eq30}) have a 
power-law localization amplitude for $p$ and a box-function localization 
amplitude for $\theta$. 

We have seen that both types of modified Husimi $\;\;\;\;\;$ functions can 
be evaluated on a classical computer with ${\cal O}(N\log(N)^2)$ 
operations. Based on the idea of the QFT which 
is closely related to the FFT if simulated 
on a classical computer, we have also implemented an efficient 
classical algorithm for the original Husimi function (\ref{eq22}) 
with the gaussian coherent states. For each value of $p_0$, we only 
consider the restricted sum such that $|p-p_0|\le 4\sqrt{N}$ and 
evaluate the matrix elements $\<\varphi(p_0,\theta_0)\,|\,\psi\>$ 
for all values of $\theta_0$ simultaneously using FFT. This provides 
also an algorithm with complexity ${\cal O}(N\log(N)^2)$ but with a 
considerably larger numerical pre-factor. However, this method does 
not allow for a ``pure'' quantum computation as it is possible 
for the two types of modified Husimi functions. 

\begin{figure}[h]
\begin{center}
\includegraphics[width=0.48\textwidth]{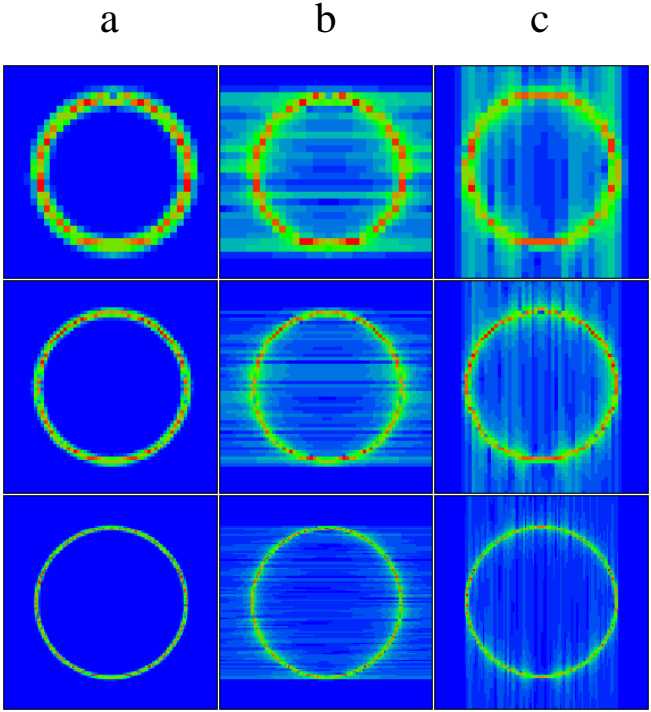}
\end{center}
\caption{\label{fig13} 
Density plots of the three different types of Husimi functions 
for the circle-state (\ref{eq31}) with $n_q=10,\ 12,\ 14$
(rows from top to bottom). 
The horizontal axis corresponds 
to $\theta_0\in[0,2\pi[$ and the vertical axis to $p_0\in[0,N[$. 
The density is minimal for blue/black
and maximal for red/white. 
Column (a) corresponds to the Husimi function (\ref{eq22}) 
with gaussian amplitude, column (b) to the momentum modified 
Husimi function (\ref{eq25}) and column (c) to the  phase
modified Husimi function (\ref{eq30}). 
}
\end{figure}

In order to compare the different Husimi functions, we consider a 
test-state defined by a circular superposition of coherent states as:
\begin{equation}
\label{eq31}
|\psi_{\rm circle}\>=
\tilde A\sum_{(p_0,\theta_0)\in\bigcirc} |\varphi(p_0,\theta_0)\>\ .
\end{equation}
Here $\tilde A$ is a normalization constant and the sum runs over a 
discrete set of points $(\theta_0,p_0)$ on a circle with center 
$(\pi,N/2)$ and relative diameter 0.7 (as compared to the size of the phase
space). 

In Fig. \ref{fig13}, we show the density plots for the three types of Husimi 
functions for this test-state with $n_q=10,\ 12,\ 14$. In all cases 
the circle picture of the density is quite well reproduced and one 
clearly sees that the circle has a finite width according to the 
widths of the coherent states due to the quantum uncertainty principle 
(see Eq. (\ref{eq24})). 
For the modified Husimi function (\ref{eq25}) (column (b)), 
one clearly observes the effect of the power-law 
decrease for the $\theta$-amplitude leading to a smearing out of 
maxima in the $\theta$-direction. The same holds for the second 
modified Husimi function (\ref{eq30}) (column (c)) 
concerning the $p$-direction. This effect is strongest for 
small values of $n_q$ and becomes smaller with increasing $n_q$. 
The effect of smearing out is not visible for the original 
Husimi function (\ref{eq22}) (column (a)) with gaussian amplitudes 
for $\theta$ and $p$. 

In order, to study the evolution of the Husimi function
for chaotic and regular regimes in the quantum tent map, we choose as 
initial condition the circle-state (\ref{eq31}). 
The semi-classical density of this state intersects quite 
well with both regular and chaotic parts of the mixed 
phase space (see Figs.~\ref{fig1},\ref{fig2}).

\begin{figure}[h]
\begin{center}
\includegraphics[width=0.48\textwidth]{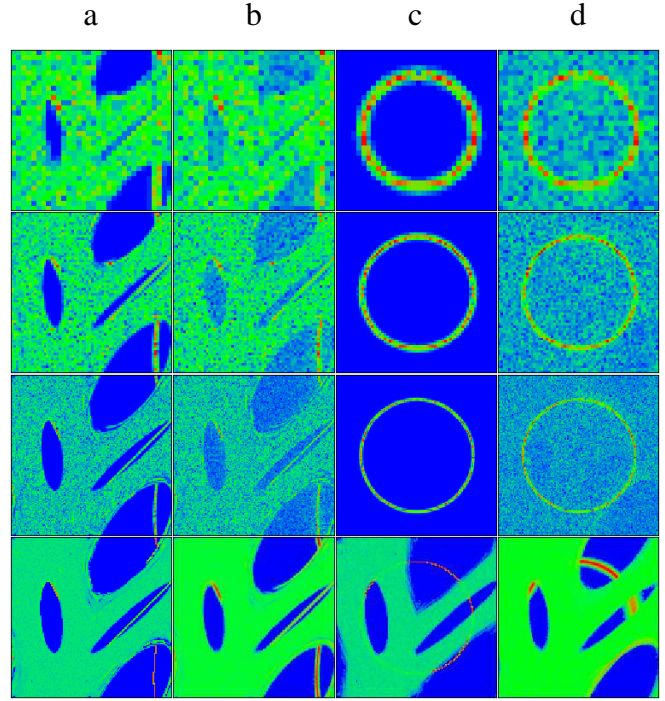}
\end{center}
\caption{\label{fig14} 
Density plot of the Husimi function (\ref{eq22}) from the 
the circle-state (\ref{eq31}) after 100 iterations with 
the quantum map with $K=kT=1.7$ and $T=2\pi/N=2\pi/2^{n_q}$ 
(columns (a) and (b)) and after 100 iterations 
in the future and 100 iterations with the inverse map in the past 
(columns (c) and (d)). The first three rows correspond to 
$n_q=10,\ 12,\ 14$ (from top to bottom)
and the last row corresponds a classical 
density plot obtained from a histogram-sampling with a box-size 
corresponding to the resolution for $n_q=14$ and with an average number 
of 100 classical trajectories per box. Columns (a) and (c) correspond 
to the exact quantum or classical maps only limited by the relative 
machine precision $10^{-16}$. Columns (b) and (d) correspond 
to the quantum computation
with random errors in quantum gates 
($\varepsilon = 0.01$) or perturbed the classical map 
perturbed by noise with 
$\varepsilon=0.01$ (see text).
}
\end{figure}

In Fig. \ref{fig14} we show the density-plots of the Husimi functions 
(defined by Eq. (\ref{eq22}) 
of the state obtained from the circle-state after 100 iterations of the 
quantum map for the cases $n_q=10,\ 12,\ 14$ (column (a)). 
We also show the state that is obtained by applying further 100 iterations 
of the inverse quantum map which should theoretically provide the 
original circle-state (column (c)). In Fig. \ref{fig14} we also 
show a density-plot for the classical map (\ref{eq3}) (with $p$ to be taken 
modulo $2\pi/T$). Here we have determined the classical trajectories 
of $100N$ random initial points on the circle. Then the density-plot has been 
calculated from a histogram with a finite box-size corresponding to the 
finite resolution of the quantum case with $n_q=14$. 

One clearly sees in column (a) that the chaotic part of the phase space is 
filled up ergodically while the piece of the circle intersecting the regular 
part of the phase space remains a connected line. 
Actually, this line rotates with a constant angular velocity 
around the center fix-point due to the local linear behavior of 
$V'(\theta)$ close to the fix-point. 

Concerning the states obtained after the back-iteration in time,
which are shown in column (c), 
one observes that the inverse 
quantum map reproduces exactly the initial state while for the classical 
map only the pieces of the circle belonging to the 
regular part of the phase space 
are reproduced. This is due to the finite machine precision (of $10^{-16}$) 
together with the exponential instability in the chaotic part of the 
phase space. We have verified that for only 25 iterations, the circle 
is well reproduced in every part of the phase space. At 50 iterations the 
classical computer round-off errors already have significant effects 
but are not sufficient to 
create a uniform distribution in the chaotic region as 
it is the case with 100 iterations shown in Fig. \ref{fig14}, last row of 
column (c). This effect is completely absent in the quantum simulation. 
The information for the phase space distribution is encoded in 
the quantum state in such a way that it is not sensible to the 
round-off errors of the classical computer simulating the quantum algorithm
for the tent map. 

To investigate this point in more detail, we have also performed a quantum 
simulation where all quantum gates are perturbed by random errors
(see Sec.~3). The effects of this noisy perturbation can be seen 
in Fig.~\ref{fig14} in the 
columns (b) (100 forward iterations of the circle-state) and 
(d) (100 forward and 100 backward iterations) where we have chosen 
$\varepsilon=0.01$. Concerning the 
quantum map, the noise reduces some-how the general quality of the 
pictures but it does not distinguish between chaotic and regular 
regions of the phase space. In particular in column (d), the circular 
density is quite well reproduced with some additional overall noise. 
Concerning the classical map, the circle-pieces in the regular region 
still remain closed lines but they acquire a finite width which increases 
in a diffusive way with the number of iterations. 
The circle-pieces in the chaotic region become 
very quickly mixed. Furthermore, it is not possible to reproduce the 
initial circle in the chaotic region due to the exponential instability 
(last row of column (d)). We have verified that this effect is already 
true for only 15 forward and 15 backward iterations if $\varepsilon=0.01$
(for the classical map the noise is introduced in the equation for 
momentum with an amplitude $ \varepsilon=0.01$, Fig.~\ref{fig14} bottom d).

\begin{figure}[h]
\begin{center}
\includegraphics[width=0.48\textwidth]{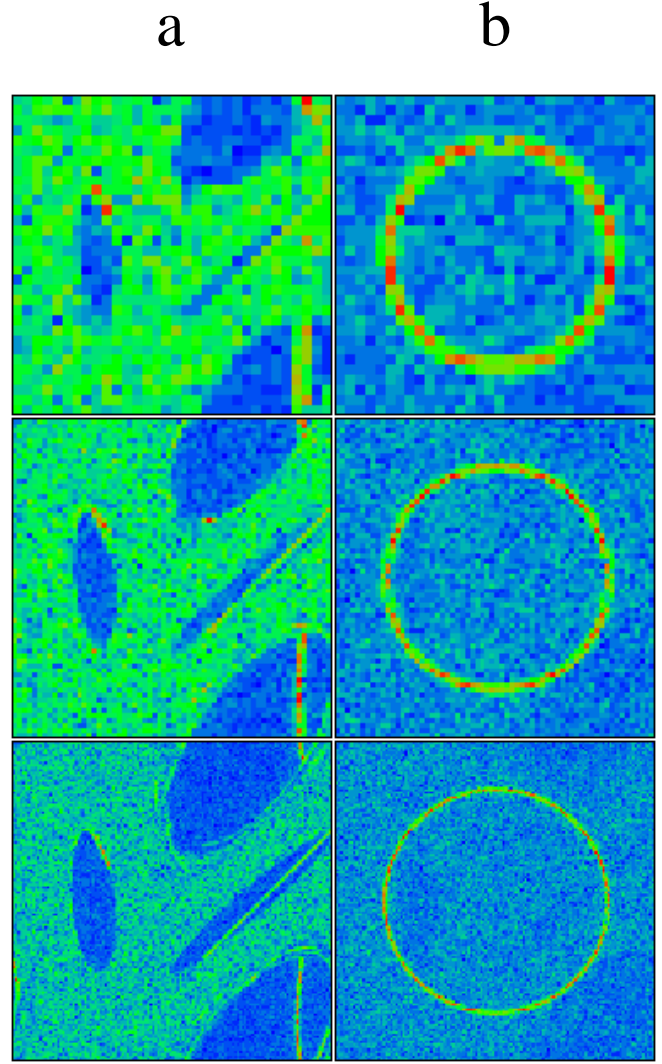}
\end{center}
\caption{\label{fig15} 
Density plot of the Husimi function (\ref{eq22}) from the 
the circle-state (\ref{eq31}) after 100 iterations with 
the quantum map (same values as in Fig. \ref{fig14}) 
perturbed by static errors (see text) 
with $\varepsilon=10^{-5}$ (column (a)) and after 100 iterations 
in the future and 100 iterations with the inverse perturbed map in the past 
(column (b)). The different rows correspond to 
$n_q=10,\ 12,\ 14$ (from top to bottom). 
}
\end{figure}

We have also studied the effects of static imperfections on 
the Husimi function evolution in the tent map.
In Fig. \ref{fig15}, we show the results of 
static imperfections with $\varepsilon=10^{-5}$ and $n_q=10,\ 12,\ 14$. The 
initial state is again the circle-state and 
column (a) corresponds to the state after 100 forward iterations 
and column (b) to the state after 100 forward and 100 backward iterations. 
The effect is quite similar to the quantum computation with random errors 
(columns (b) and (d) of Fig. \ref{fig14}). The general quality of the 
pictures is reduced and there is no distinction between regular and 
chaotic part of the phase space. Again, in column (b) the circular 
density is quite well reproduced with some additional overall noise.
We should note that  the static imperfections of 
strength $\varepsilon =10^{-5}$ give perturbations in the Husimi function 
which are comparable with those in the case of random errors at
$\varepsilon = 0.01$. This shows that the static imperfections 
perturb the quantum computations in a stronger way
comparing to random errors.

\begin{figure}[h]
\begin{center}
\includegraphics[width=0.48\textwidth]{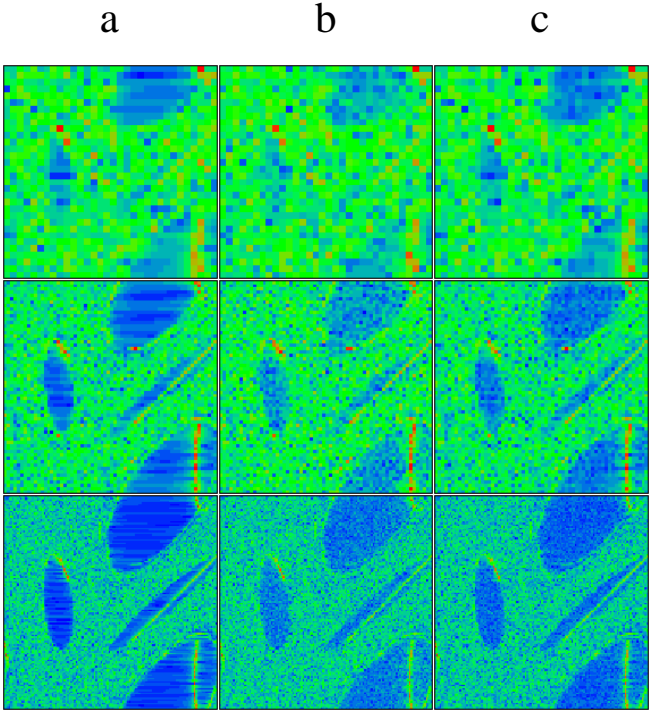}
\end{center}
\caption{\label{fig16} 
Density plot of the modified Husimi function (\ref{eq25}) from the 
the circle-state (\ref{eq31}) after 100 iterations with 
the quantum map (same values as in Fig. \ref{fig3}). 
Column (a) corresponds to the exact quantum map, column (b) to the 
map with random errors in quantum gates
with $\varepsilon=0.01$ and column (c) 
to the quantum map simulated 
with static imperfections with $\varepsilon=10^{-5}$. 
The different rows correspond to $n_q=10,\ 12,\ 14$ (from top to bottom). 
}
\end{figure}

Finally, we show in Fig. \ref{fig16} the modified Husimi functions 
(\ref{eq25}) after 100 iterations applied to the initial circle-state 
again for the three cases $n_q=10,\ 12,\ 14$. Column (a) shows the 
exact simulation, (b) the case of random errors ($\varepsilon=0.01$) and 
(c) the quantum map with static imperfections ($\varepsilon=10^{-5}$). 
We note that the smearing-out effect discussed at the beginning 
of this Section  (see Fig.~\ref{fig13}) 
is well visible for the case of the exact simulation, while 
it is not visible at all for the cases with random errors 
or static imperfections. 
Therefore, the utilization of the modified Husimi function seems to be 
quite well justified in these cases. 

\section{Conclusion}
\label{sec8}

The results obtained in this paper give a universal description of
fidelity decay in quantum algorithms 
simulating complex dynamics on a realistic quantum computer with
static imperfections. This decay is given by Eq.~\ref{rmt_eq27}
which determines the time scale $t_f$ of reliable quantum
computation with fidelity $f > 0.9$. According to  Eq.~\ref{rmt_eq27}
\begin{equation}
\label{eqfinal1}
t_f \approx t_c/10  = 1/(10\varepsilon^2 n_q n_g^2) \; ; \;
N_g \approx 1/(10 \varepsilon^2 n_q n_g)
\end{equation}
for $t_{\rm H} > t_c$ so that 
$\varepsilon > \varepsilon_{ch}= 2^{-n_q/2}/(n_g \sqrt{n_q} )$.
Here, $N_g=t_fn_g$ is the total number of gates which can be performed with
fidelity $f>0.9$.
In this regime the static errors act in a way similar to random noise errors
even if their effect is stronger due to coherent
accumulation of static errors inside a certain interval
of the algorithm (one map iteration for the tent map).
Indeed, for random errors in quantum gates the relation (\ref{eq4.1})
gives
\begin{equation}
\label{eqfinal2}
t_f \approx t_r/10  \approx  5/(\varepsilon^2 n_g) \; ; \;
N_g \approx 5/ \varepsilon^2 .
\end{equation}
We note that (\ref{eqfinal2}) is in agreement with the result
obtained for random errors in a very different 
quantum algorithm \cite{wavelet} and hence it is generic.
Even if the dependence of $N_g$ on $\varepsilon$ in Eqs. (\ref{eqfinal1}),
(\ref{eqfinal2}) is the same, the dependence on $n_q$ is rather different.
This difference should play an important role for
the quantum error correction codes 
which allow to perform
the fault-tolerant quantum computation for the random error rate
 $p_r \sim \epsilon^2 < 10^{-4}$ (see {\it e.g.} 
\cite{steane,chuang,gottesman,aharonov,steane1}).
The fact that for random errors $N_g$ is independent of $n_q$
while for static imperfections $N_g$ drops strongly with 
$n_q$ should significantly decrease the threshold for fault-tolerant
quantum computation in presence of static imperfections. 
 
For $t_c < t_{\rm H}$ or 
$\varepsilon < \varepsilon_{ch} = 2^{-n_q/2}/(n_g \sqrt{n_q})$
the time scale $t_f$ is given by the relation
\begin{equation}
\label{eqfinal3}
 t_f \approx 0.2 \sqrt {t_c t_{\rm H}} 
\approx 2^{n_q/2} /(5 \varepsilon n_g \sqrt{n_q}) \; ; \;
N_g \approx 2^{n_q/2} /(5 \varepsilon \sqrt{n_q}).
\end{equation}
In this regime the effect of static imperfections is absolutely different from
random noise errors. This regime may be dominant for up to 10 - 15 qubits.
However, in the limit of large $n_q \gg 10$ it appears only in the limit
of very small static imperfections and should not be very important
for quantum computers with few tens of qubits. 
The transition from the regime (\ref{eqfinal1}) to regime (\ref{eqfinal3})
takes place for 
\begin{equation}
\label{eqfinal4}
\varepsilon > \varepsilon_{ch} = 2^{-n_q/2}/(n_g \sqrt{n_q} )\; .
\end{equation}
From the physical point of view this border can be interpreted
as the quantum chaos border above which the static imperfections 
start to mix the energy levels of ideal quantum algorithm.
The fact that this border drops exponentially with the number
of qubits $n_q$ has been discussed in \cite{benentiqcb}
for a quantum algorithm for complex dynamics.
Above $\varepsilon_{ch}$ the effect of static imperfections becomes
somewhat similar to random errors.

The results (\ref{eqfinal1})
and (\ref{eqfinal3}) for the time scales of reliable quantum computation
are based on the RMT approach and are universal for algorithms
which simulate a complex dynamics, {\it e.g.} an evolution
in the regime of quantum chaos. 
However, it is important to keep in mind that there are other 
types of algorithms where the evolution is rather regular,
{\it e.g.} the Grover algorithm or integrable dynamics.
In such cases the asymptotic dependence of $t_{\rm H}$ on $n_q$
should be studied in more detail. It is not excluded that
in such cases $t_{\rm H}$ grows with $n_q$ very slowly 
(see Fig.~\ref{fig11}) or even may be independent of $n_q$.
In such situations the static imperfections will generate
a very significant reduction of the time scale 
of reliable quantum computation. In a sense our RMT result 
(\ref{rmt_eq27}) gives the weakest form of fidelity decay in a 
realistic quantum computer with $n_q$ qubits 
since the reduction of the chaotic component $\sigma$ accelerates 
this decay. 

The universal regime for fidelity decay in quantum computations established 
in this paper can also find other applications. For example it can 
appear in the decay of spin echo in interacting spin systems.

This work was supported in part by the EC IST-FET project
EDIQIP and the NSA and ARDA under ARO contract No. DAAD19-01-1-0553
and by the French goverment grant ACI Nanosciences-Nanotechnologies LOGIQUANT.
We thank CalMiP at Toulouse and  IDRIS at Orsay 
for access to their supercomputers.

Upon completion of this manuscript a recent preprint of T. Gorin, T.~Prosen, 
and T.~H.~Seligman \cite{gorin} came to our attention where the relation 
between fidelity decay and two-level form factor has been established 
for an abstract Hamiltonian model with continuous time evolution and 
a perturbation given by an invariant random matrix. However, in this 
work only the regime of a small perturbation strength $\lambda\ll 1$ 
$(\lambda \sim \sqrt{t_H/t_c})$ is studied 
which translates to $t_{\rm H} \ll t_c$ with the dominant gaussian decay.



\begin{thebibliography}{99}
\bibitem{josza} A.~Ekert and R.~Josza, Rev. of Mod. Phys. {\bf 68}, 733 (1996).
\bibitem{steane}  A.~Steane, Rep. Prog. Phys. {\bf 61}, 117 (1998).
\bibitem{chuang} M.A.~Nielsen and I.L.~Chuang {\it Quantum Computation and 
       Quantum Information}, Cambridge Univ. Press, Cambridge (2000).
\bibitem{divi} D.~P.~Di~Vincenzo, Science {\bf 270}, 255 (1995).
\bibitem{shor} P.W.~Shor, in {\it Proc. 35th Annual Symposium
       on Foundation of Computer Science}, Ed. S.Goldwasser (IEEE Computer
       Society, Los Alamitos, CA, 1994), p.124
\bibitem{grover} L.~K.~Grover, Phys. Rev. Lett. {\bf 79}, 325 (1997).
\bibitem{lloyd} S.~Lloyd, Science {\bf 273}, 1073 (1996).
\bibitem{ortiz} G.~Ortiz, J.E.~Gubernatis, E.~Knill, and R.Laflamme,
       Phys. Rev. A {\bf 64}, 22319 (2001).
\bibitem{chirikov} B.V.Chirikov, F.M.Izrailev and D.L.Shepelyansky, 
     Sov. Scient. Rev. C (Gordon \& Bridge) {\bf 2}, 209 (1981);
     Physica D {\bf 33}, 77 (1988).
\bibitem{izrailev} F.M.Izrailev, Phys. Rep. {\bf 196}, 299 (1990).
\bibitem{schack} R.~Schack, Phys. Rev. A {\bf 57}, 1634 (1998).
\bibitem{georgeotkr} B.~Georgeot, and D.L.~Shepelyansky, Phys. Rev. Lett. 
       {\bf 86}, 2890 (2001).
\bibitem{benentist} G.~Benenti, G.~Casati, S.~Montangero, and 
       D.L.~Shepelyansky, Phys. Rev. Lett. {\bf 87}, 227901 (2001).
\bibitem{well} A.D.~Chepelianskii and D.L.~Shepelyansky, 
       Phys. Rev. A {\bf 66}, 054301 (2002).
\bibitem{pomeransky} A.A.~Pomeransky and D.L.~Shepelyansky,
       Phys. Rev. A (to appear), quant-ph/0306203.
\bibitem{cat}  B.~Georgeot and D.L.~Shepelyansky,
       Phys. Rev. Lett. {\bf 86}, 5393 (2001); {\bf ibid.}
       {\bf 88}, 219802 (2002).
\bibitem{stratt} M.~Terraneo, B.~Georgeot and  D.L.~Shepelyansky,
       Eur. Phys. J. D {\bf 22}, 127 (2003).
\bibitem{zurek} W.H.~Zurek, Rev. Mod. Phys. {\bf 75}, 715 (2003).
\bibitem{georgeot} B.Georgeot and D.L.Shepelyansky, Phys. Rev. E {\bf 62}, 
       3504 (2000); {\bf 62}, 6366 (2000).
\bibitem{flambaum} V.V.~Flambaum, Aust. J. Phys. {\bf 53}, 489 (2000).
\bibitem{berman} G. P. Berman, F. Borgonovi, F. M. Izrailev, 
     and V. I. Tsifrinovich, Phys. Rev. E {\bf 64}, 056226 (2001).
\bibitem{benentieu} G.~Benenti, G.~Casati and D.L.~Shepelyansky, 
        Eur. Phys. J. D {\bf 17}, 265 (2001).
\bibitem{braun} D.~Braun, Phys. Rev. A {\bf 65}, 042317 (2002).
\bibitem{peres} A.~Peres, Phys. Rev. A {\bf 30}, 1610 (1984). 
\bibitem{dls1983} D.L.~Shepelyansky, Physica D {\bf 8}, 208  (1983).
\bibitem{casati1986} G.~Casati, B.V.~Chirikov, I.~Guarneri
        and D.L.~Shepelyansky, Phys. Rev. Lett. {\bf 56}, 2437 (1986).
\bibitem{pastawski} R.A.~Jalabert and H.M.~Pastawski, Phys. Rev. Lett.
        {\bf 86}, 2490 (2001).
\bibitem{beenakker} P.Jacquod, P.G.Silvestrov and C.W.J.~Beenakker, 
        Phys. Rev. E  {\bf 64}, 055203 (2001).
\bibitem{veble} G.~Veble and T.~Prosen, Phys. Rev. Lett. {\bf 92}, 
	034101 (2004).
\bibitem{como} G.~Benenti and G.~Casati, Phys. Rev. E {\bf 66}, 066205 (2002).
\bibitem{prosen} T.~Prosen and M.~Znidaric, J. Phys. A {\bf 34}, L681 (2001);
	J. Phys. A 35, 1455 (2002); T.~Prosen, T.H.~Seligman and M.~Znidaric, 
	Prog. Theor. Phys. Supp. {\bf 150}, 200 (2003). 
\bibitem{cohen} T.~Kottos and D.Cohen, Europhys. Lett. {\bf 61}, 431 (2003).
\bibitem{cerruti} N.R.~Cerruti and S.~Tomsovic,  Phys. Rev. Lett. {\bf 88}, 
	054103 (2002); J. Phys. A 36, 3451 (2003). 
\bibitem{adamov} Y.~Adamov, I.V.~Gornyi and A.D.~Mirlin, Phys. Rev. E 
	{\bf 67}, 056217 (2003).
\bibitem{paz} C.~Miguel, J.P.~Paz and W.H.~Zurek, Phys. Rev. Lett.
        {\bf 78}, 3971 (1997).
\bibitem{wavelet} M.~Terraneo and D.L.~Shepelyansky, Phys. Rev. Lett.
        {\bf 90}, 257902 (2003).
\bibitem{bettelli} S.~Bettelli, quant-ph/0310152.
\bibitem{dyson} F.~J.~Dyson, J. Math. Phys. {\bf 3}, 140 (1962).
\bibitem{mehta} M.L.~Mehta, {\it Random Matrices} (Academic, New York, 1991).
\bibitem{guhr} T.~Guhr, A.~Mueller-Groeling and H.A.~Weidenmueller,
        Phys. Rep. {\bf 299}, 189 (1998).
\bibitem{wigner}  E.~Wigner, Phys. Rev. {\bf 40}, 749 (1932).
\bibitem{husimi} S.-J.~Chang and K.-J.~Shi, Phys. Rev. A {\bf 34}, 7 (1986).
\bibitem{chuang1} L.M.K.Vandersypen, M. Steffen, G. Breyta,
       C.S. Yannoni, M.H. Sherwood, and I.L. Chuang,
       Nature {\bf 414}, 883 (2001).
\bibitem{cory} Y.S.~Weinstein, S.~Lloyd, J.~Emerson, and D.G.~Cory,
       Phys. Rev. Lett. {\bf 89}, 157902 (2002).
\bibitem{emerson} J.~Emerson, Y.S.~Weinstein, S.~Lloyd and D.G.~Cory,
       Phys. Rev. Lett. {\bf 89}, 284102 (2002).
\bibitem{chirikov79} B.V.~Chirikov, Phys. Rep. {\bf 52}, 263 (1979). 
\bibitem{tent} S.~Bullett, Com. Math. Phys. {\bf 107}, 241 (1986).
\bibitem{vecheslavov} V.V.~Vecheslavov, Zh. Eksp. Teor. Fiz. {\bf 119},
       853 (2001) (nlin.CD/0005048); V.V.~Vecheslavov and B.V.~Chirikov,
       Zh. Eksp. Teor. Fiz. {\bf 120}, 740 (2001) (nlin.CD/0202017).
\bibitem{khmelnitski} B.A.~Muzykantskii and D.E.~Khmelnitskii, Phys. Rev. B
       {\bf 51}, 5481 (1995).
\bibitem{mirlin} A.D.~Mirlin, Phys. Rep. {\bf 326}, 259 (2000).
\bibitem{maspero} G.~Casati, G.~Maspero and D.L.~Shepelyansky,
       Phys. Rev. E {\bf 56}, R6233 (1997).
\bibitem{sokolov} D.V.~Savin and V.V.~Sokolov, Phys. Rev. E {\bf 56},
       R4911 (1997).
\bibitem{frahm} K.M.~Frahm, Phys. Rev. E {\bf 56}, R6237 (1997).
\bibitem{gottesman} D.~Gottesman, Phys. Rev. A {\bf 57}, 127 (1998).
\bibitem{aharonov} D.~Aharonov and M.~Ben-Or, quant-ph/9906129.
\bibitem{steane1}  A.~Steane, quant-ph/0207119.
\bibitem{benentiqcb} G.~Benenti, G.~Casati, S.~Montangero and 
        D.L.~Shepelyansky, Eur. Phys. J. D {\bf 20}, 293 (2002).
\bibitem{altshuler}  B.~L.~Altshuler and B.~I.~Shklovskii, Zh. Eksp. Teor. 
	Fiz. {\bf 91}, 220 (1986) [Sov. Phys. JETP {\bf 64}, 127 (1986)].
\bibitem{gorin} T.~Gorin, T.~Prosen, and T.~H.~Seligman, preprint nlin. 
	CD/0311022v1.


\end{thebibliography}
\end{document}